\documentclass[letterpaper,11pt]{article}
\pdfoutput=1

\usepackage[margin=1in,letterpaper]{geometry}
\usepackage{latexsym}
\usepackage[T1]{fontenc}
\usepackage[utf8]{inputenc}
\usepackage[english]{babel}
\usepackage{cite}
\usepackage{amsmath,amssymb}
\usepackage{mathtools}
\usepackage{empheq}
\usepackage{physics}
\usepackage{bm}
\usepackage{slashed}
\usepackage{latexsym}
\usepackage{srcltx}
\usepackage{graphicx,epsfig}
\usepackage{subfigure}
\usepackage{tikz-cd}
\usepackage{float}

\usepackage{booktabs}
\usepackage{tabularx}
\usepackage{dcolumn}

\usepackage{setspace}
\usepackage{enumitem,amssymb}
\usepackage{ulem}
\usepackage[nottoc]{tocbibind}

\usepackage{hyperref}
\usepackage{xurl}
\usepackage{url}
\usepackage{listings}
\usepackage{color}

\usepackage{comment}
\usepackage{pdfpages}
\usepackage[most]{tcolorbox}
\usepackage{pifont}

\definecolor{mpurple}{rgb}{0.44, 0.16, 0.39}
\definecolor{myblue}{rgb}{0,0,0.8}
\definecolor{darkgreen}{rgb}{0,0.5,0}
\definecolor{purple}{rgb}{1,0,1}
\definecolor{codegreen}{rgb}{0,0.6,0}
\definecolor{codegray}{rgb}{0.5,0.5,0.5}
\definecolor{codepurple}{rgb}{0.58,0,0.82}
\definecolor{backcolour}{rgb}{0.95,0.95,0.92}

\numberwithin{equation}{section}

\lstdefinestyle{mystyle}{
	backgroundcolor=\color{backcolour},
	commentstyle=\color{codegreen},
	keywordstyle=\color{magenta},
	numberstyle=\tiny\color{codegray},
	stringstyle=\color{codepurple},
	basicstyle=\ttfamily\footnotesize,
	breakatwhitespace=false,
	breaklines=true,
	captionpos=b,
	keepspaces=true,
	showspaces=false,
	showstringspaces=false,
	showtabs=false,
	tabsize=2
}
\newcount\Comments
\newcommand{\kibitz}[2]{\ifnum\Comments=1\textcolor{#1}{#2}\fi}

\newlist{todolist}{itemize}{2}
\setlist[todolist]{label=$\square$}

\newcommand{\be}{\begin{equation}}
	\newcommand{\ee}{\end{equation}}

\newcommand{\br}{\begin{eqnarray}}
	\newcommand{\er}{\end{eqnarray}}

\newcommand{\al}{\alpha}

\newcommand{\g}{\gamma}

\newcommand{\eps}{\epsilon}

\newcommand{\la}{\lambda}

\newcommand{\pa}{\partial}

\newcommand{\lie}{\mathcal{G}}

\begin{document}
	\vspace*{1cm}
\noindent

\vskip 1 cm
\begin{center}
	{\Large\bf New soliton solutions for Chen-Lee-Liu and Burgers hierarchies and its Bäcklund transformations\footnote{This paper is dedicated to the memory of Abraham Hirsz Zimerman,  1928-2025, a dear friend, mentor and long-term collaborator.}}
\end{center}
\normalsize
\vskip 1cm

\begin{center}
	{{Y. F. Adans}$^{a,b}$,
	{H. Aratyn}$^{c}$,
	{C. P. Constantinidis}$^{d}$, 
	{J. F. Gomes}$^{a}$, 
	{G. V. Lobo}$^{a}$, 
	and 
	{T. C. Santiago}$^{a}$.
	\\[.5cm]
	
	\par \vskip .1in \noindent
	$^{a}$\emph{Universidade Estadual Paulista (Unesp), Instituto de Física Teórica (IFT), São Paulo,
		Rua Dr. Bento Teobaldo Ferraz 271, 01140-070, São Paulo, SP, Brasil}\\[0.3cm]
		
	$^{b}$ \emph{School of Mathematics \& Hamilton Mathematics Institute, 
	Trinity College Dublin, Ireland}\\[0.3cm]
	
	$^{c}$\emph{Department of Physics, University of Illinois Chicago,\\ 845 W. Taylor St., 60607-7059, Chicago, IL, USA.}\\[0.3cm]

	$^{d}$\emph{Universidade Federal do Espirito Santo, Depto. de Física,\\ Av. Fernando Ferrari, 514., CEP 29075-900, Vitoria, ES, Brasil.}\\[0.5cm]
\texttt{ysla.franca@unesp.br}, \texttt{aratyn@uic.edu}, \texttt{clisthenis.constantinidis@ufes.br},  \texttt{francisco.gomes@unesp.br}, 
\texttt{gabriel.lobo@unesp.br} ,\texttt{t.santiago@unesp.br}. \\
	\vskip 2cm}
	
\end{center}
		
		\begin{abstract}
			
			Positive and negative flows of the
			Chen-Lee-Liu model and its various
			reductions, including Burgers
			hierarchy, are formulated within the
			framework of Riemann-Hilbert-Birkhoff
			decomposition with the constant grade
			two generator. Two classes of vacua,
			namely zero vacuum and constant non-zero vacuum
			can be realized	
			within a centerless Heisenberg algebra. The tau
			functions for soliton solutions are
			obtained by a dressing method and
			vertex operators are constructed for
			both types of vacua. We are able to
			select and classify the soliton
			solutions in terms of the type of
			vertices involved. {A judicious choice of vertices yields in a closed form a particular set of multi soliton solutions for the Burgers hierarchy.} We {develop and analyze}  a class of gauge-Bäcklund transformations  that generate further multi soliton solutions from those obtained by dressing method by letting them  interact with various integrable defects.
			
		\end{abstract}
		
		
		
%
%
%
%
%
		
	
	

	\section{Introduction}
	\label{introduction}
	
	Integrable hierarchies are often realized as
	two-dimensional field theories that allow an infinite
	number of conservation laws which, in turn 
	ensure stability of soliton solutions. A
	crucial ingredient in constructing such integrable
	hierarchies with an underlying affine algebraic
	structure is its gradation 
	\cite{drinfeld_lie_1985,de_groot_generalized_1992}.
	The flow equations are conveniently obtained in terms of a zero
	curvature representation,
	\begin{align}\label{zcc}
		\comm{\pa_x + A_x(\phi)}{\pa_{t_ {\pm N}}+ A _{t_{\pm N}}(\phi)} = 0,
		\qquad
		A_x, A_{t_{\pm N}} \in \hat{\lie}, 
		\qquad  N \in \mathbb{N^*}. 
	\end{align}	
	Notice that \eqref{zcc} generate a series of flows associated to graded Heisenberg algebra elements. The construction is well-known for positive flows. More recently negative flows have been incorporated \cite{aratyn_complex_2000, gomes_negative_2009, adans_negative_2023, aratyn_generalized_2025} and shown to generate new interesting symmetries \cite{adler_negative_2024, adler_3d_2024, kolesnikov_negative_2025}. 
	
	There are many {ansatzes} for
	constructing the auxiliary, field
	dependent, two-dimensional gauge
	potentials (Lax operators) $A_x(\phi
	)$ and $A_{t_{\pm N}}(\phi )$.  
	Many well-known examples, as mKdV and
	AKNS,  
	involve hierarchies classified according
	to the {\it grading} {of} the affine
	Lie algebra $\hat{\lie} = \sum_{i\in
		\mathbb{Z}}\hat{\lie}_i$ and a choice
	of {\it grade one semi-simple}
	generator $E^{(1)}$.  
	
	A systematic approach for constructing the Lax operators can be formulated in terms of the Riemann-Hilbert-Birkhoff (RHB) decomposition (see for instance \cite{aratyn_integrable_2003}). The underlying algebraic framework is very powerful and allows for the systematic construction of soliton solutions from representation theory.  The dressing method constructs soliton solutions employing a gauge transformation to map the Lax operators from a particular	vacuum solution, $A_x(\phi _{vac})$ and $A_{t_{\pm N}}(\phi _{vac})$ into a non-trivial configuration, $A_x(\phi)$ and $A_{t_{\pm N}}(\phi )$. 
	
	There are however examples involving higher grade semi-simple elements,	$E^{(a)}, \; a>1$	\cite{ferreira_affine_1996} and	presenting a variety of non-trivial	boundary conditions with different vacuum solutions	\cite{gomes_negative_2009, gomes_nonvanishing_2012}.
	
	In this paper  we follow   a proposal \cite{aratyn_generalized_2025} for {the} generalized Riemann-Hilbert-Birkhoff (g-RHB) decomposition formula  that includes both, {\it higher grading  semi-simple elements} and  a {\it variety of  non-trivial vacuum configurations}. The condition to  encompass  different  vacuum solutions requires  the existence of  Heisenberg  sub-algebras. In fact, Heisenberg sub-algebras classify  the possible  boundary conditions.  Define the generalized Baker-Akhiezer function (g-BA), 
	\begin{align}\label{B-A}
		\Psi_a = e^ {-\sum_{N}{ \left(\eps^ {(aN)}{t_{ N}} + \eps^ {(-aN)}{t_{ -N}}\right)}}.
	\end{align}
	Here,   $\eps^{(\pm aN)}, \; a\in \mathbb{N^*}$ are vacuum parameters dependent generators that satisfy  a centerless Heisenberg  algebra $[\eps^{(aM)},\; \eps^{(aN)}]=0$. Notice that $\Psi_a $ 
	displays explicit  space-time information  (where $ t_1 \equiv x$).

	The simplest example corresponds to
	the mKdV hierarchy with $a=1$.  The
	Lax operators acting on vacuum were
	constructed in
	\cite{gomes_negative_2009,
		aratyn_generalized_2025} and were shown to
	generate one-parameter deformed
	Heisenberg algebras for {\it positive
		odd} and {\it negative even} flows.

	In this paper we engage the g-RHB decomposition (\ref{g-rh}) and (\ref{B-A}) with $a=2$ to formulate the	Chen-Lee-Liu (CLL) hierarchy, and	construct its soliton solutions in terms of different possible vacuum solutions and their reductions to Burgers hierarchy.
	
	In section
	\ref{sec.gRHB.decomposition} we
	discuss 
	the construction of
	the positive and negative flows for
	the CLL hierarchy in terms of various
	Heisenberg sub-algebras, each
	describing different possible vacuum
	solutions.  
	
	In section \ref{sec.flows} the various reductions to {\it heat } and {\it Burgers}
	equations are discussed.  The systematic construction of soliton
	solutions is presented explicitly in Section 4. The algebraic structure provides an elegant construction for soliton solutions. An important element introduced by the Kyoto School approach \cite{jimbo_solitons_1983} is the associated vertex operators which correspond to eigenvectors of the Heisenberg algebras. The associated eigenvalues encode the space-time dependence for the soliton solutions. It is interesting to note that these vertices may depend upon vacuum parameters and henceforth provide a new class of soliton solutions.	
	
	The dressing method
	employed here follows directly from the g-RHB decomposition and
	implies gauge transforming the g-BA $\Psi_a$ with vacuum information
	to some non-trivial solution $\Phi =\Theta_{+ }\Psi_a= \Theta_{-}
	\Psi_a g$. This is accomplished by the construction of a pair of
	vertex operators, namely $V_{\pm }$.  The solutions are then
	classified into {\it class A}, when powers of only one of the
	vertices, either $V_{+}$ or $V_{-}$ are considered and as a
	consequence, one of the {fields} remains constant (non vanishing). Such
	structure uncovers the underlying Burgers hierarchy associated to
	class A solutions and the dressing method generates, in closed form,
	the $n-$soliton solution for the entire Burgers hierarchy. The
	second, {\it {class B}}, is obtained when powers of the product $V_+
	V_-$ {are} considered and both fields are shown to be non trivial.  
	
	In section 5 we construct a gauge-Bäcklund transformation as a
	generalization of the dressing method, where two non-trivial
	solutions are connected by gauge transformation.  The Bäcklund
	transformation {is} shown to describe integrable defects \cite{corrigan_new_2009, corrigan_type_2018} 
	since it describes the connection between two solutions at a specific space position.  We then discuss explicit examples of possible integrable defects.  The key
	ingredient is an ansatz involving three consecutive graded terms with
	the virtue to accommodate two non-trivial soliton configurations. 
	
	In section 6 we discuss in detail the two classes of Bäcklund
	solutions. Since class A {contains} powers of a single vertex operator
	and one of the fields, either $r$ or $s$, {remains} constant for all flow
	equations.  The CLL hierarchy then reduces to the Burgers hierarchy and so does
	the corresponding Bäcklund transformation.  We therefore discuss the
	scattering and transition of one-soliton and two-solitons solutions
	for Burgers hierarchy.
	
	Next we consider, in section 7, class B of
	Bäcklund solutions composed of powers of mixed
	vertices. We discuss the scattering of
	one-soliton and the transition of one to
	two-soliton solutions for the CLL hierarchy.

	\section{The Generalized Riemann-Hilbert-Birkhoff (g-RHB) Decomposition}	
	\label{sec.gRHB.decomposition}

	Consider the generalized Baker-Akhiezer  function 
	(g-BA) (\ref{B-A}).   The connection with integrable 
	hierarchies  is established with the  
	identification of Heisenberg generators  with  
	vacuum configuration,
	\begin{equation}
		\begin{split}\label{dec}
			A_{t_N}^ {vac} &=  A_{t_N}(\phi_{vac}) = E^ {(aN)}	 + D_{vac}^  {(aN-1)} + \cdots + D_{vac}^  {(0)}\equiv \eps^ {(aN)},
			\\[2.5mm]
			A_{t_{-N}}^ {vac} &= A_{t_{-N}}(\phi_{vac})= E^ {(-aN)}	 + D_{vac}^  {(-aN-1)} + \cdots + D_{vac}^{(-1)} \equiv \eps^ {(-aN)},
		\end{split}
		\qquad
		\begin{split}
			D_{vac}^  {(i)} \in \hat{\lie_i}
		\end{split}
	\end{equation}
	where $A_{t_{1}} \equiv A_x$. For zero ($\phi_{vac}=0 $) or nonzero vacuum ($\phi_{vac} = \phi_0$) configurations,  the zero curvature representation (\ref{zcc}) yields  an important (centerless) Heisenberg  algebra which may depend upon  complex  parameters, namely ($\phi_{0}$) \cite{aratyn_integrable_2003, aratyn_generalized_2025},
	\begin{align}\label{heis}
		\comm{A_x^ {vac}}{ A_{t_{\pm N}}^ {vac}}
		=
		\comm{A_x(\phi_{vac})}{A_{t_{\pm N}}(\phi_{vac})}
		= 0.
	\end{align} 
	
	In order to   derive a construction  of the two dimensional  Lax operators $A_x$ and $ A_{t_ N} $ consider   the following  g-RHB decomposition
	\begin{align}\label{g-rh}
		\Theta (t) = \Psi_a (t)\; g \; \Psi_a^ {-1} (t) = \Theta_-^ {-1}(t) \Theta_+(t)\,
	\end{align}
	where $g$ is an  arbitrary constant group element and 
	\begin{align} \label{fr} 
		\Theta_-(t) = \tilde B \prod_{k=1}^ {\infty} e^ {- \theta^{(-k)}}, 
		\qquad 
		\Theta_+(t) = \tilde B \; B\;\prod_{k=1}^{\infty} e^ { \theta^ {(k)}}, 
		\quad
		B=e^ {\theta^ {(0)}} , 
		\quad \tilde B = e^ {\tilde \theta_0},
		\qquad \theta^ {(k)}\in \hat{\lie_k}. 
	\end{align}

	Notice that (\ref{g-rh})  \textit{does not depend} upon $\tilde B$. Here $\tilde B$ represents a {\it gauge freedom } and can be chosen 
	for convenience as $\tilde B = B^ {-c}, \;\; 0\leq c\leq 1$ such 
	that allows one to  reshoufle the  zero grade component to   be contained partially 
	within the positive, $\Theta_+ \rightarrow  B ^{-c}\Theta_+$  or negative , $\Theta_-\rightarrow  B^ {-c} \Theta_-$ graded subgroups as shown in (\ref{fr}).

	The flow structure ($t=t_{N}$) of integrable hierarchies is 
	determined by a decomposition of an affine algebra into graded
	subspaces, $\hat{\lie} = \sum_{i\in \mathbb{Z}} \hat{\lie}_i$, and its corresponding
	decomposition of $A_x(\phi)$ and $ A_{t_{ N}}(\phi)$, as 
	discussed in detail in the next sections.  
	
	In particular, in \cite{aratyn_generalized_2025} it was shown that integrable
	hierarchies depend upon two distinct structures, \textbf{i)} constant semisimple operators  of (higher) grade $a \in \mathbb{N^*}$, $ E^ {(a)}$, 	and \textbf{ii)} nonzero constant vacuum parameters defined from the
	Lax operators in vacuum, $A_x^ {vac}(\phi_0) $ and $ A_{t_{\pm N}}^ {vac}(\phi_0)$ . 
	\begin{subequations}
		\begin{align}
			A_x &=\Theta_{\pm}A_x^ {vac} \Theta_{\pm}^ {-1}- \left(\pa_x \Theta_{\pm} \right)\Theta_{\pm}^ {-1} =( \Theta_{-} \eps^ {(a)} \Theta_{-}^{-1} )_{\geq} -\left(\pa_{x}B^{-c}\right)B^c=E^{(a)}+\sum_{i=0}^{a-1} A_{i} 
			\label{ax}
			\\[2mm]
			A_{t_{ N}} &= \Theta_{\pm}A_{t_{ N}} ^ {vac} \Theta_{\pm}^ {-1}- \left(\pa_{t_{ N}}  \Theta_{\pm}\right) \Theta_{\pm}^ {-1}=( \Theta_{-} \eps^ {(aN)} \Theta_{-}^{-1})_{\geq}-\left(\pa_{t_N}B^{-c}\right)B^c=E^{(aN)}+\sum_{i=0}^ {aN-1} D^{(i)}
			\label{amtpos}
			\\[2mm]
			A_{t_{ -N}} &=\Theta_{\pm}A_{t_{ -N}} ^ {vac} \Theta_{\pm}^ {-1}- \left(\pa_{t_{ -N}}  \Theta_{\pm}\right) \Theta_{\pm}^ {-1}=( \Theta_{+} \eps^ {(-aN)} \Theta_{+}^ {-1} )_{<}-\left(\pa_{t_{-N}}B^{-c}\right)B^c= E^{(-aN)}+\sum_{i=0}^ {aN-1} D^{(-i)} 
			\label{amt}
		\end{align}
	\end{subequations}
	Notice that $\Theta_{\pm}$ are identified  with the dressing matrices  mapping the vacuum $A^ {vac}_{\mu} (\phi_0)$ to some non-trivial configuration, $A_{\mu} (\phi)$. In fact  the g-RHB decomposition (\ref{g-rh}) is the basis of the dressing method where non-trivial solutions are constructed  from a specific vacuum configuration \cite{babelon_dressing_1992, babelon_affine_1993, ferreira_tau-functions_1997}.
	An important ingredient  here is the construction of vertex operators which correspond to eigenvalues and eigenstates of the Heisenberg algebra denoted by $\eps^ {(aN)}$ encoded within the generalized Baker-Akhiezer function (\ref{B-A}), and henceforth depend upon the vacuum through the vacuum parameters $\phi_0$. 
	
	On the other hand, equations \eqref{ax}-\eqref{amt} naturally generalizes to the idea of connecting two distinct configurations by gauge transformation, i.e., 
	\begin{align}\label{zz}
		A_{\mu}(\phi) = U^{-1} A_{\mu}(\psi) U- \pa_{\mu}U U^{-1}, 
		\qquad 
		(\mu = x \; \text{or} \; t_{\pm N}),
	\end{align}
	where $U(\phi, \psi)$ that depends of field configurations and eqn. \eqref{zz} generate the {\it gauge-Bäcklund transformation}.
	
	In fact, this is the key idea  in constructing  Bäcklund as gauge  transformation acting on the  two dimensional   potentials such that  the zero curvature and therefore, the equations of motion  
	remain unchanged.  It is important to note that  Bäcklund transformation  connects  two  distinct solutions of the same equation.  
	In particular, eqn. (\ref{ax})-(\ref{amt}) represent 
	the case where $\psi $ denotes the vacuum
	configuration. Such framework was proposed and
	employed to describe integrable defects in the sense
	that the two solutions are interpolated by a defect
	\cite{bowcock_classically_2004,
		corrigan_jump-defects_2006,
		caudrelier_systematic_2008}.


	\section{Lax pair for the Chen-Lee-Liu (CLL) flows}
	\label{sec.flows}
	
	Consider the  loop-algebra $ L(\mathcal{G})=\{h^{(n)}, E_{ \alpha}^{(n)}, E_{-\alpha}^{(n)} \}$, endowed with the principal gradation, see \ref{app.algebra}. The grading operator $Q_{p}=\tfrac{1}{2}h^{(0)}+2\hat{d}$ decomposes the algebra $L(\mathcal{G}) = \sum_{i\in \mathbb{Z}}\lie_i$ into graded subspaces:
	\begin{align*}
		\mathcal{G}_{2m}=\left \{ h^{(m)}\right \}, \qquad \mathcal{G}_{2m+1}=\left \{ E^{(m)}_{\alpha}, E^{(m+1)}_{-\alpha}\right \}, \qquad n,m \in \mathbb{Z},
	\end{align*}
	of grade  $2m$ and  $2m+1$, respectively. 
	A second decomposition  of  $L(\mathcal{G})$ into Kernel $\mathcal{K}$ and its complement $\mathcal{M}$:
	\begin{align*}
		\mathcal{K}=\left\{ h^{(n)} \right\}, \qquad \mathcal{M}=\left\{ E_{\alpha}^{(n)}, E_{-\alpha}^{(n)} \right\},
	\end{align*}
	is generated by  a constant, grade two generator, $E^{(2)}=\frac{1}{2}h^{(1)} \in \lie_2 $.
	The kernel and its complement satisfy the following relations:
	\begin{align*}
		\comm{\mathcal{K}}{\mathcal{K}} \subset \mathcal{K}, 
		\quad \quad 
		\comm{\mathcal{K}}{\mathcal{M}}\subset \mathcal{M}, 
		\quad \quad 
		\comm{\mathcal{M}}{\mathcal{M}} \subset \mathcal{K}.
	\end{align*}
	The above algebraic structure underlies the  Chen-Lee-Liu (CLL) hierarchy, which can be derived from the spatial Lax operator 
	with $a=2$ and $c=\tfrac{1}{2}$ in \eqref{ax}, \cite{aratyn_generalized_2025, chen_k_2002, chen_integrability_1979, zhang_discrete_2022}:
	\begin{align}\label{Ax1}
		A_{x}= E^{(2)} + r  E_{\alpha}^{(0)} + s  E_{-\alpha}^{(1)} -\frac{1}{2} r  s  h^{(0)},
	\end{align}
	where  $r=r\left(x,t_{\pm N}\right)$ and
	$s=s\left(x,t_{\pm N}\right)$ are fields of
	the theory associated  to positive (or
	negative) flows $t_{ N}$ (or $t_{-N}$), with  $N \in \mathbb{N}^*$. 
	
	The flow equations associated to the  Lax operator \eqref{Ax1} are obtained  by solving the zero curvature equation 
	\footnote{In the case of flow  $t_1$, the solution is trivial: $A_{t_1}=A_x$.}
	\begin{equation}\label{cn}
		\comm{\pa_x+A_x}{\pa_{t_{\pm N}}+A_{t_{\pm N}}}=\partial_xA_{t_{\pm N}}-\partial_{t_{\pm N}}A_x+\comm{A_x}{A_{t_{\pm N}}}=0,
	\end{equation}
	where $A_{t_{\pm N}}$ is the temporal  Lax potential associated to a given $t_{\pm N}$. For positive and negative sub-hierarchies their structure is respectively given by
	\begin{equation}\label{At+}
		A_{t_{N}}=E^{(2N)}+\sum_{i=0}^{2N-1}D^{(i)},
	\end{equation}
	and
	\begin{equation}
		\label{At-}
		A_{t_{-N}}=E^{(-2N)}+\sum_{i=0}^{2N-1}D^{(-i)},
	\end{equation}
	where
	\begin{equation*}
		E^{(\pm 2N)} =\tfrac{1}{2}h^{(\pm N)},
		\qquad
		D^{(2j)}=a_{2j,\pm N}h^{(j)},
		\qquad
		D^{(2j+1)}=b_{2j+1,\pm N}E_{\alpha}^{(j)}+c_{2j+1,\pm N}E_{-\alpha}^{(j+1)}
	\end{equation*}
	and  $j \in \mathbb{Z}$ , $a_{2j,\pm N}$, $b_{2j+1,\pm N}$ and  $c_{2j+1,\pm N}$ are functions of  $x$ and  $t_{\pm N}$, to be determined.	
	The flow equations are therefore obtained by solving (\ref{cn}) for  either (\ref{At+}) or  (\ref{At-}).
	
	\subsection{Positive flows}
	
	For  the positive sub-hierarchy  we find from   (\ref{Ax1}) and  (\ref{At+}) in  the zero curvature  equation (\ref{cn}),
	\begin{equation}
		\label{cn+}
		\comm{\pa_x+E^{(2)}+A_1+A_0}{\pa_{t_{N}}+E^{(2N)}+D^{(2N-1)}+D^{(2N-2)}+\cdots+D^{(1)}+D^{(0)}}=0,
	\end{equation}
	where $E^{(2)}=\frac{1}{2}h^{(1)}$, $A_1=rE_{\alpha}^{(0)}+sE_{-\alpha}^{(1)}$ and  $A_0=-\frac{1}{2}rs h^{(0)}$. The general structure of $L(\mathcal{G})$   decomposes (\ref{cn}) into graded subspaces, i.e.,
	\begin{equation}
		\begin{aligned} \label{cn++}
			\comm{E^{(2)}}{E^{(2N)}}=0, & \\
			\comm{A_{1}}{E^{(2N)}}+\comm{E^{(2)}}{D^{(2N-1)}}=0, & \\
			\comm{A_{0}}{E^{(2N)}}+\comm{A_{1}}{D^{(2N-1)}}+\comm{E^{(2)}}{D^{(2N-2)}}=0, & \\
			\partial_x D^{(2N-1)}+\comm{A_{0}}{D^{(2N-1)}}+\comm{A_{1}}{D^{(2N-2)}}+\comm{E^{(2)}}{D^{(2N-3)}}=0, & \\
			\vdots \hspace{5mm} & \\
			\partial_x D^{(2)}+\comm{A_{0}}{D^{(2)}}+\comm{A_{1}}{D^{(1)}}+\comm{E^{(2)}}{D^{(0)}}=0, & \\
			\partial_{t_N}A_{1}-\partial_x D^{(1)}-\comm{A_{0}}{D^{(1)}}-\comm{A_{1}}{D^{(0)}}=0, & \\
			\partial_{t_N}A_{0}-\partial_x D^{(0)}-\comm{A_{0}}{D^{(0)}}=0. &
		\end{aligned}
	\end{equation}
	We start solving  from   the highest grade equation, namely,   $2N+1$, in order to determine the coefficients  $b_{2N-1,N}$ and $c_{2N-1,N}$ in terms of fields  $r$ e $s$.   We solve recursively all  equations until  
	the grade one component obtaining in this process the equations of motion:
	\begin{align}
		\partial_{t_N}r=\partial_xb_{1,N}-r\left( 2a_{0, N}+b_{1, N}s\right), 
		\qquad 
		\partial_{t_N}s=\partial_xc_{1,N}+s\left( 2a_{0, N}+c_{1, N}r\right). \label{pos-b}
	\end{align}
	Solving (\ref{cn++}) for the first few  flows we find for $N=2$:
	\begin{equation}
		\begin{aligned}
			A_{t_{2}}&=\frac{1}{2}h^{(2)}+rE_{\alpha}^{(1)}+sE_{-\alpha}^{(2)}-rsh^{(1)}-\left(  r^2s+\partial_xr\right)E_{\alpha}^{(0)}+\left( -rs^2+\partial_xs\right)E_{-\alpha}^{(1)}+\\
			&+\frac{1}{2}\left(r^2s^2-r\partial_xs+s\partial_xr \right) h^{(0)}.
		\end{aligned}
	\end{equation}
	Repeating the procedure for  $N=3$ yields:
	\begin{equation*}
		\begin{aligned}
			A_{t_{3}}&=\frac{1}{2}h^{(3)}+rE_{\alpha}^{(2)}+sE_{-\alpha}^{(3)}-rsh^{(2)}-\left(  r^2s+\partial_xr\right)E_{\alpha}^{(1)}+\left( -rs^2+\partial_xs\right)E_{-\alpha}^{(2)}+\\
			&+\left(r^2s^2-r\partial_xs+s\partial_xr \right) h^{(1)}+\left[r\left(r^2s^2+3s\partial_xr-r\partial_xs \right)+\partial_x^2r  \right]E_{\alpha}^{(0)}+\\
			&+\left[s\left(r^2s^2-3r\partial_xs+s\partial_xr \right)+\partial_x^2s \right]E_{-\alpha}^{(1)}+\\
			&-\frac{1}{2}\left[ r^3s^3-\left(\partial_xr\right)\left(\partial_xs\right)+r\left(-3rs\partial_xs+\partial_x^2s\right)+s\left(3rs\partial_xr+\partial_x^2r\right) \right]h^{(0)}.
		\end{aligned}
	\end{equation*}
	and for  $N=4$:
	\begin{equation*}
		\begin{aligned}
			A_{t_{4}}&=\frac{1}{2}h^{(4)}+rE_{\alpha}^{(3)}+sE_{-\alpha}^{(4)}-rsh^{(3)}-\left(  r^2s+\partial_xr\right)E_{\alpha}^{(2)}+\left( -rs^2+\partial_xs\right)E_{-\alpha}^{(1)}+\\
			&+\left(r^2s^2-r\partial_xs+s\partial_xr \right) h^{(2)}+\left[r\left(r^2s^2+3s\partial_xr-r\partial_xs \right)+\partial_x^2r  \right]E_{\alpha}^{(1)}+\\
			&+\left[s\left(r^2s^2-3r\partial_xs+s\partial_xr \right)+\partial_x^2s \right]E_{-\alpha}^{(2)}+\\
			&-\left[r^3s^3-\left(\partial_xr\right)\left(\partial_xs\right)+r\left(-3rs\partial_xs+\partial_x^2s\right)+s\left(3rs\partial_xr+\partial_x^2r\right) \right]h^{(1)}+\\
			&-\left\{ r\Bigl[r^3s^3-\left(\partial_xr\right)\left(\partial_xs\right)+r\left(-3rs\partial_xs+\partial_x^2s\right)+2s\left(3rs\partial_xr+2\partial_x^2r \right)\Bigr]+3s\left(\partial_xr\right)^2+\partial_x^3r\right\}E_{\alpha}^{(0)}+\\
			&+\left\{-s\Bigl[r^3s^3-\left(\partial_xr\right)\left(\partial_xs\right)+s\left( 3rs\partial_xr+\partial_x^2r\right)+2r\left( -3rs\partial_xs+2\partial_x^2s\right)\Bigr]-3r\left(\partial_xs\right)^2+\partial_x^3s \right\}E_{-\alpha}^{(1)}+\\
			&+\frac{1}{2}\Bigl\{ r^4s^4 -4rs\left(\partial_xr\right)\left(\partial_xs\right)+\left(\partial_xr\right)\left(\partial_x^2s\right)-\left(\partial_xs\right)\left(\partial_x^2r\right)+\Bigr. \\
			&-\Bigl. r\left[6r^2s^2\partial_xs-3r\left(\partial_xs\right)^2 +4rs\partial_x^2s+\partial_x^3s\right]+s\left[6r^2s^2\partial_xr+3s\left(\partial_xr\right)^2 -4rs\partial_x^2r+\partial_x^3r\right]\Bigr\}h^{(0)}.
		\end{aligned}
	\end{equation*}
	leading respectively to the following time evolution equations,
	\begin{equation}
		\begin{aligned}
			\partial_{t_2}r&=-\partial_x^2r-2rs\partial_xr, 
			\\
			\partial_{t_2}s&=\partial_x^2s-2rs\partial_xs,
		\end{aligned}
	\end{equation}
	
	\begin{equation}
		\begin{aligned}
			\partial_{t_3}r&=\partial_x^3r+3r^2s^2\partial_xr+3s\partial_x\left(r\partial_xr\right),
			\\
			\partial_{t_3}s&=\partial_x^3s+3r^2s^2\partial_xs-3r\partial_x\left(s\partial_xs\right), 
		\end{aligned}
	\end{equation}
	
	\begin{equation}
		\begin{aligned}
			\partial_{t_4}r&=-\partial_x^4r-4r^3s^3\partial_xr-6s^2\partial_x\left( r^2\partial_xr\right)-4s\partial_x\left(r\partial_x^2r\right)-6s\left(\partial_xr\right)\left(\partial_x^2r\right)-2\partial_x\left[ r\left(\partial_xr\right)\left(\partial_xs\right)\right],
			\\
			\partial_{t_4}s&=\partial_x^4s-4r^3s^3\partial_xs+6r^2\partial_x\left( s^2\partial_xs\right)-4r\partial_x\left(s\partial_x^2s\right)-6r\left(\partial_xs\right)\left(\partial_x^2s\right)-2\partial_x\left[ s\left(\partial_xr\right)\left(\partial_xs\right)\right] .
		\end{aligned} 
	\end{equation}
	The above equations of motion admit two classes of vacuum solutions, 
	$ i)$  zero vacuum, i.e. $r=s=0$ and $ ii)$ strictly nonzero  constant vacuum, 
	$r=r_0\neq 0, \,  s=s_0 \neq 0$ solutions.
	Consider now the  general vacuum configuration for the Lax operators	
	\begin{equation}
		\begin{aligned} \label{vac-pos}
			A_{t_1}^{vac}&=\Sigma^{(2)},
			\\[2mm]
			A_{t_2}^{vac}&=\Sigma^{(4)}-br_0s_0\Sigma^{(2)},
			\\[2mm]
			A_{t_3}^{vac}&=\Sigma^{(6)}-br_0s_0\Sigma^{(4)}+br_0^2s_0^2\Sigma^{(2)},
			\\[2mm]
			A_{t_4}^{vac}&=\Sigma^{(8)}-br_0s_0\Sigma^{(6)}+br_0^2s_0^2\Sigma^{(4)}-br_0^3s_0^3\Sigma^{(2)}, 
		\end{aligned}
	\end{equation}
	where 
	\begin{equation} \label{vac-ppos}
		\Sigma^{(2N)}\equiv\frac{1}{2}\left(h^{(N)}-br_0s_0 h^{(N-1)}\right)+br_0E_{\alpha}^{(N-1)}+bs_0E_{-\alpha}^{(N)}.
	\end{equation}
	and the parameter  $b$ is used to classify the two classes of vacua namely,  $b=0$ for zero vacuum and $b=1$ for  nonzero constant  vacuum solutions\footnote{For b=1, mixed vacuum configurations $(r,s)=(r_0,0)$ and $(r,s)=(0,s_0)$ can also be considered.}.  It therefore follows that $\comm{\Sigma^{(2)}}{\Sigma^{(2N)}}=0$ and henceforth,
	\begin{align*}
		\comm{\Sigma^{(2M)}}{\Sigma^{(2N)}}=0, \qquad M,N=1,2, \cdots
	\end{align*}
	for either  $b=0$ or $b=1$.

	\subsection{Negative flows}
	In order to construct the negative  flows we insert the Lax pair from equations (\ref{Ax1}) and  (\ref{At-}) into the zero curvature  equation  (\ref{cn}),
	\begin{equation}
		\label{cn-}
		\comm{\pa_x+E^{(2)}+A_1+A_0}  
		{\pa_{t_{-N}}+E^{(-2N)}+D^{(-2N+1)}+D^{(-2N+2)}+\cdots+D^{(-1)}+D^{(0)}}=0\,
	\end{equation}
	and decompose it into the graded subspaces,
	\begin{equation}
		\begin{aligned} \label{cn--}
			\comm{A_0}{E^{(-2N)}}&=0, \\
			\partial_xD^{(-2N+1)}+\comm{A_1}{E^{(-2N)}}+\comm{A_0}{D^{(-2N+1)}}&=0, \\
			\partial_xD^{(-2N+2)}+\comm{E^{(2)}}{E^{(-2N)}}+\comm{A_1}{D^{(-2N+1)}}+\comm{A_0}{D^{(-2N+2)}}&=0, \\
			\vdots \hspace{5mm} & \\
			\partial_xD^{(-1)}+\comm{E^{(2)}}{D^{(-3)}}+\comm{A_1}{D^{(-2)}}+\comm{A_0}{D^{(-1)}}&=0, \\
			\partial_{t_{-N}}A_0-\partial_xD^{(0)}-\comm{E^{(2)}}{D^{(-2)}}-\comm{A_1}{D^{(-1)}}+\comm{A_0}{D^{(0)}}&=0, \\
			\partial_{t_{-N}}A_1-\comm{E^{(2)}}{D^{(-1)}}-\comm{A_1}{D^{(0)}}&=0, \\
			\comm{E^{(2)}}{D^{(0)}}&=0. \\
		\end{aligned}
	\end{equation}
	The lowest grade component  $-2N+1$ now determines  the coefficients $b_{-2N+1,N}$ and  $c_{-2N+1,N}$ in terms of fields  $r$ and $s$. 
	The procedure follows recursively  until we reach the time evolution equation, 
	\begin{align}
		\partial_{t_{-N}}r=b_{-1,-N}-2a_{0,-N}r, \qquad \partial_{t_{-N}}s=-c_{-1,-N}+2a_{0,-N}s. \label{neg-b}
	\end{align}
	Solving for the first few flows, we find in terms of  new variables, 
	\begin{equation}
		R\equiv\partial_x^{-1}\left(r e^{-J}\right), \quad 
		S\equiv\partial_x^{-1}\left(s e^{J}\right), \quad 
		J=\partial_x^{-1}\left( r s \right).
	\end{equation}
	where we employed  more compact notation of $\partial_x^{-1}f=\int^{x} f(y)\, dy$.  We therefore find, 
	for $N=1$:
	\begin{equation}
		\begin{aligned}
			A_{t_{-1}}=\frac{1}{2}h^{(-1)}+e^{J} R E_{\alpha}^{(-1)}-e^{-J} S E_{-\alpha}^{(0)}+\frac{1}{2} R S h^{(0)}.
		\end{aligned}
	\end{equation}
	For  $N=2$ we find:
	\begin{equation}
		\begin{aligned}
			A_{t_{-2}}&=\frac{1}{2}h^{(-2)}+e^{J} R E_{\alpha}^{(-2)}-e^{-J} S E_{-\alpha}^{(-1)}+R S h^{(-1)}-e^{J}\partial_x^{-1}\left( R-2 R S\partial_x R\right) E_{\alpha}^{(-1)}+ \\
			&-e^{-J}\partial_x^{-1}\left( S+2 R S\partial_x S\right)E_{-\alpha}^{(0)}-\frac{1}{2}\left[\left(R S\right)^2-R \partial_x^{-1}\left( S+2 R S\partial_xs\right)+ S \partial_x^{-1}\left( R-2 R S\partial_x R\right)\right] h^{(0)},
		\end{aligned}
	\end{equation}
	and  for $N=3$:
	\begin{equation}
		\begin{aligned}
			A_{t_{-3}}&=\frac{1}{2}h^{(-3)}+e^{J} R E_{\alpha}^{(-3)}-e^{-J} S E_{-\alpha}^{(-2)}+R S h^{(-2)}-e^{J}\partial_x^{-1}\left( R-2 R S\partial_x R\right) E_{\alpha}^{(-2)}+\\
			&-e^{-J}\partial_x^{-1}\left( S+2 R S\partial_x S\right)E_{-\alpha}^{(-1)}-\left[\left(R S\right)^2-R \partial_x^{-1}\left( S+2 R S\partial_x S\right)+ S \partial_x^{-1}\left( R-2 R S\partial_x R\right)\right] h^{(-1)}+\\
			&+e^{J}\partial_x^{-1}\Bigl\{ \left(1-2 S\partial_x R \right)\partial_x^{-1}\left( R-2 R S\partial_x R\right)-2\partial_x R \left[\left(R S\right)^2-R\partial_x^{-1}\left( S+2 R S \partial_x S\right)\right]\Bigr\}E_{\alpha}^{(-1)}+\\
			&-e^{-J}\partial_x^{-1}\Bigl\{ \left( 1 +2 R \partial_x S \right)\partial_x^{-1}\left( S+2 R S\partial_x S\right)-2\partial_x S\left[\left( R S\right)^2+S\partial_x^{-1}\left(R-2 R S\partial_x R\right)\right]\Bigr\}E_{-\alpha}^{(0)}+\\
			&-\frac{1}{2}\Bigl\{ -2\left(R S\right)^3+\partial_x^{-1}\left(R-2 R S\partial_x R\right)\partial_x^{-1}\left( S+2 R S\partial_x S\right)+\\
			&-2 R S\left[R\partial_x^{-1}\left( S+2 R S \partial_x S\right)+S\partial_x^{-1}\left( R-2 R S \partial_x R \right)\right] +\nonumber \\
			&-R\partial_x^{-1}\Bigl[\left(1-2 S \partial_x R \right)\partial_x^{-1}\left( R-2 R S\partial_x R\right)-2\left(R S \right)^2\partial_x R+2 R\left(\partial_x R\right)\partial_x^{-1}\left( S+2 R S\partial_x S\right)\Bigr]+\nonumber\\
			&-S\partial_x^{-1}\Bigl[\left(1-2 S \partial_x R \right)\partial_x^{-1}\left( R-2 R S\partial_x R\right)-2\left(R S \right)^2\partial_x R+2 R\left(\partial_x R\right)\partial_x^{-1}\left( S+2 R S\partial_x S\right)\Bigr]\Bigr\}h^{(0)}.
		\end{aligned}
	\end{equation} 
	yielding respectively  the following time evolution equations,  
	\begin{align}
		&\partial_{t_{-1}}r=Re^{J}-rRS, \\
		&\partial_{t_{-1}}s=Se^{-J}+sRS,
	\end{align}
	\begin{align}
		&\partial_{t_{-2}}r=-e^{J}\partial_x^{-1}\left( R-2 R S\partial_x R\right)+r\left[\left(R S\right)^2-R \partial_x^{-1}\left( S+2 R S\partial_xS\right)+ S \partial_x^{-1}\left( R-2 R S\partial_x R\right)\right], \\
		&\partial_{t_{-2}}s=e^{-J}\partial_x^{-1}\left( S+2 R S\partial_x S\right)-s\left[\left(R S\right)^2-R \partial_x^{-1}\left( S+2 R S\partial_xS\right)+ S \partial_x^{-1}\left( R-2 R S\partial_x R\right)\right],
	\end{align}
	\begin{align}
		\partial_{t_{-3}}r&=e^{J}\partial_x^{-1}\Bigl\{ \left(1-2 S\partial_x R \right)\partial_x^{-1}\left( R-2 R S\partial_x R\right)-2\partial_x R \left[\left(R S\right)^2-R\partial_x^{-1}\left( S+2 R S \partial_x S\right)\right]\Bigr\}+\\
		&+r\Bigl\{ -2\left(R S\right)^3+\partial_x^{-1}\left(R-2 R S\partial_x R\right)\partial_x^{-1}\left( S+2 R S\partial_x S\right)+\nonumber \\
		&-2 R S\left[R\partial_x^{-1}\left( S+2 R S \partial_x S\right)+S\partial_x^{-1}\left( R-2 R S \partial_x R \right)\right] +\nonumber \\
		&-R\partial_x^{-1}\Bigl[\left(1-2 S \partial_x R \right)\partial_x^{-1}\left( R-2 R S\partial_x R\right)-2\left(R S \right)^2\partial_x R+2 R\left(\partial_x R\right)\partial_x^{-1}\left( S+2 R S\partial_x S\right)\Bigr]+\nonumber\\
		&-S\partial_x^{-1}\Bigl[\left(1-2 S \partial_x R \right)\partial_x^{-1}\left( R-2 R S\partial_x R\right)-2\left(R S \right)^2\partial_x R+2 R\left(\partial_x R\right)\partial_x^{-1}\left( S+2 R S\partial_x S\right)\Bigr]\Bigr\}, \nonumber
		\\
		\partial_{t_{-3}}s&=e^{-J}\partial_x^{-1}\Bigl\{ \left( 1 +2 R \partial_x S \right)\partial_x^{-1}\left( S+2 R S\partial_x S\right)-2\partial_x S\left[\left( R S\right)^2+S\partial_x^{-1}\left(R-2 R S\partial_x R\right)\right]\Bigr\}+\\
		&-s\Bigl\{ -2\left(R S\right)^3+\partial_x^{-1}\left(R-2 R S\partial_x R\right)\partial_x^{-1}\left( S+2 R S\partial_x S\right)+\nonumber \\
		&-2 R S\left[R\partial_x^{-1}\left( S+2 R S \partial_x S\right)+S\partial_x^{-1}\left( R-2 R S \partial_x R \right)\right] +\nonumber \\
		&-R\partial_x^{-1}\Bigl[\left(1-2 S \partial_x R \right)\partial_x^{-1}\left( R-2 R S\partial_x R\right)-2\left(R S \right)^2\partial_x R+2 R\left(\partial_x R\right)\partial_x^{-1}\left( S+2 R S\partial_x S\right)\Bigr]+\nonumber\\
		&-S\partial_x^{-1}\Bigl[\left(1-2 S \partial_x R \right)\partial_x^{-1}\left( R-2 R S\partial_x R\right)-2\left(R S \right)^2\partial_x R+2 R\left(\partial_x R\right)\partial_x^{-1}\left( S+2 R S\partial_x S\right)\Bigr]\Bigr\}. \nonumber
	\end{align}
	Notice that  all  the above equations  admit both zero  vacuum ($r=0,\, s=0$)  or  nonzero constant vacuum solutions,  ($r=r_0,\, s=s_0$).
	For both cases  we  define the  vacuum  configuration Lax operators  for the negative sub-hierarchy.
	Considering the limits   ($r\, \to \, 0,\, s\, \to \,0,\, R\, \to \, 0,\, S\, \to \,0$) or ($r\, \to \,r_0,\, s\, \to \,s_0,\, R\, \to \, -\frac{1}{s_0}e^{-r_0s_0x},\, S\, \to \,\frac{1}{r_0}e^{r_0s_0x}$) we find,
	\begin{equation}
		\begin{aligned} \label{vac-neg}
			A_{t_{-1}}^{vac}&=\Upsilon^{(-2)},
			\\[2mm]
			A_{t_{-2}}^{vac}&=\Upsilon^{(-4)}-\frac{b}{r_0s_0}\Upsilon^{(-2)},
			\\[2mm]
			A_{t_{-3}}^{vac}&=\Upsilon^{(-6)}-\frac{b}{r_0s_0}\Upsilon^{(-4)}+\frac{b}{r_0^2s_0^2}\Upsilon^{(-2)}, 
		\end{aligned}
	\end{equation}
	where
	\begin{equation} \label{vac-nneg}
		\Upsilon^{(-2N)}\equiv\frac{1}{2}\left(h^{(-N)}-\frac{b}{r_0s_0}h^{(-N+1)} \right)-\frac{b}{s_0  }E_{\alpha}^{(-N)}-\frac{b}{r_0 } E_{-\alpha}^{(-N+1)}.
	\end{equation}
	satisfying the centerless Heisenberg  algebra 
	\begin{align}
		\comm{\Upsilon^{(-2M)}}{\Upsilon^{(-2N)}}=0, \qquad M,N \in \mathbb{N},
	\end{align}
	for  $b=0$ and $b=1$. Notice that the zero vacuum limit is obtained by taking $b=0$ in the relations \eqref{vac-neg} and \eqref{vac-nneg}.
	
	We should like to point out that, from \eqref{vac-ppos} and \eqref{vac-nneg} with $b=0$, the positive and negative grade Heisenberg generators, $\Sigma^{(2M)}=h^{(M)}$ and $ \Upsilon^{(-2N)}=h^{(-N)}$ commute, i.e., $[\Sigma^ {(2M)}, \Upsilon^ {(-2N)} ]=0$, for any $M, N \in \mathbb{N}$.
	Also, for $b=1$, clearly,
	\begin{align}
		r_0 s_0  {\Upsilon ^{(-2(N-1))}} = \frac{1}{2}\left(r_0 s_0 h^{(-N)}-h^{(-N+1)} \right)-r_0 E_{\alpha}^{(-N)}-s_0 E_{-\alpha}^{(-N+1)} =   -\Sigma^{(-2N)}
	\end{align}
	and henceforth 
	\begin{align*}
		[\Sigma^ {(2M)}, \Upsilon^ {(-2N)} ]=0, 
		\qquad M,N \in \mathbb{N}
	\end{align*}
	This stresses the fact that there are two hierarchies of positive and negative commuting flows . One, the usual  CLL hierarchy associated to $b=0$  (zero vacuum ) and a new  hierarchy associated to $b=1$ (constant non-zero vacuum){\footnote{ Similar situation appears  within the mKdV case where a zero vacuum  ($b=0$) consist of positive and negative odd flows and  a second hierarchy  admitting  constant non-zero  vacuum ($b=1$) consist of positive odd and negative even flows (see  \cite{adans_negative_2023}).}}.

	\section{CLL Reductions} 
	\label{sec.reductions}
	Several interesting reductions  can be obtained  from  
	CLL hierarchy by making use of zero and  constant nonzero vacuum solutions (see Table \ref{tab:reducoesCLL}).
	\begin{table}[H]
		\centering
		\begin{tabular}{l @{\hspace{1cm}} l @{\hspace{1cm}} c}
			\toprule
			\textbf{Limit} & \textbf{Field} & \textbf{Flows}
			\\
			\midrule
			$r\to 0,\; s\to \phi$ & $\phi=\phi\left(x,t_{\pm N}\right)$ &
			\begin{tabular}{c}
				\(\partial_{t_{ N}}\phi = \partial_x^{ N} \phi\) 
				\\[2.5mm]
				\(\partial_{t_{-N}}\phi = \partial_x^{- N} \phi\) 
			\end{tabular}
			\\
			\midrule
			$r \to \psi,\; s\to 0$ & $\psi=\psi\left(x,t_{\pm N}\right)$ &
			\begin{tabular}{c}
				\(\partial_{t_{ N}}\psi = \left( -1\right)^{N+1} \partial_x^{ N} \psi\)
				\\[2.5mm]
				\(\partial_{t_{-N}}\psi = \left( -1\right)^{N+1} \partial_x^{- N} \psi\)
			\end{tabular}
			\\
			\midrule
			$r\to r_0,\; s\to w$ & $w=w\left(x,t_{\pm N}\right)$ &
			\begin{tabular}{c}
				\(\partial_{t_N}w= -\dfrac{1}{r_0}\partial_x\left[ e^{r_0\partial_x^{-1}w}\left( \partial_x^Ne^{-r_0\partial_x^{-1}w} \right) \right]\)
				\\[4mm]
				\(\partial_{t_{-N}}w= \dfrac{1}{r_0}\partial_x\left[ e^{r_0\partial_x^{-1}w}\left( \partial_x^{-N}e^{-r_0\partial_x^{-1}w} \right) \right]\)
			\end{tabular}
			\\
			\midrule
			$r \to u,\; s\to s_0$ & $u=u\left(x,t_{\pm N}\right)$ &
			\begin{tabular}{c}
				\(\partial_{t_N}u= \dfrac{1}{s_0}\left( -1\right)^{N+1}\partial_x \left[ e^{-s_0\partial_x^{-1}u}\left( \partial_x^Ne^{s_0\partial_x^{-1}u} \right) \right] \)
				\\[4mm]
				\(\partial_{t_{-N}}u= -\dfrac{1}{s_0} \left( -1\right)^{N+1} \partial_x\left[ e^{-s_0\partial_x^{-1}u}\left( \partial_x^{-N}e^{s_0\partial_x^{-1}u} \right) \right]\)
			\end{tabular}
			\\
			\bottomrule
		\end{tabular}
		\caption{Immediate reductions of the CLL hierarchy: the limits $(r\to0,\, s=\phi)$ or $(r\to\psi,\, s=0)$ yield the heat equation for $\phi$ (or $\psi$), while $(r\to r_0,\, s=w)$ or $(r=u,\, s\to s_0)$ with fixed nonzero constants $r_0$ and $s_0$ lead to the Burgers equation. The factor $(-1)^{N+1}$ can be absorbed through $t_{\pm N}\to t'_{\pm N}=t_{\pm N}/(-1)^{N+1}$, while $r_0$ and $s_0$ can be removed by the rescaling $w\to r_0 w$ and $u\to s_0 u$, showing that the models for $\phi$ and $\psi$ are equivalent, as are those for $w$ and $u$ when $u=-w$.}
		\label{tab:reducoesCLL}
	\end{table}
	
	\subsection{Burgers hierarchy}
	
	Considering the CLL hierarchy with one of the fields constrained to a constant (say $r=r_0$, see Table \ref{tab:reducoesCLL}) we obtain from (\ref{pos-b}) the positive Burgers hierarchy (\ref{wpN0}) with the positive  fluxes of the  Burgers hierarchy written in a compact closed  form \cite{kudryashov_exact_2009}:
	\begin{equation}
		\label{wpN0}
		\partial_{t_N'} w= \alpha_{N} \partial_x\left(\partial_x-r_0 w \right)^{N-1} w,
	\end{equation}
	where $t_N'=\alpha_N t_N$ and  $\alpha_N$ is an arbitrary constant. Explicitly, the first few flows can be identified to the Burgers equation for $t=t_2$, originally derived by Bateman in 1915 \cite{bateman_recent_1915} and later popularized by Burgers \cite{burgers_mathematical_1948}, 
	\begin{equation}
		\label{eqB}
		\partial_{t_2'} w= \alpha_{2}\left( \partial_x^2 w- 2r_0w\partial_x w\right),
	\end{equation}
	and the  Sharma–Tasso–Olver, derived in  \cite{sharma_connection_1977, olver_evolution_1977},
	\begin{equation}
		\label{eqSTO}
		\partial_{t_3'} w=\alpha_{3}\left[\partial_x^3 w +3r_0^2w^2\partial_x w- 3r_0\partial_x\left(w\partial_x w\right)\right]. 
	\end{equation}
	Moreover the same limiting procedure in
	(\ref{neg-b}) yields, in a closed form a new sub-hierarchy which we are dubbing {\it negative Burgers hierarchy}.
	
	The positive and negative  Burgers sub-hierarchies are  given
	in the closed form as,
	\begin{equation}
		\label{wpN}
		\partial_{t_N'} w=-\frac{\alpha_N}{r_0}\partial_x\left[ e^{r_0\partial_x^{-1}w}\left( \partial_x^Ne^{-r_0\partial_x^{-1}w} \right) \right],
	\end{equation}
	and 	
	
	\begin{equation}
		\label{wnN}
		\partial_{t_{-N}'} w=\frac{\alpha_{-N}}{r_0}\partial_x\left[ e^{r_0\partial_x^{-1}w}\left( \partial_x^{-N}e^{-r_0\partial_x^{-1}w} \right) \right],
	\end{equation}
	where $\alpha_N$ is an arbitrary constant. 
	Both cases  only admit {\it nonzero constant vacuum} solutions,  $w=w_0\neq 0$. 
	Explicitly, the first  two  flow equations  for the negative  sub-hierarchy  are:
	\begin{equation}
		\label{eq11}
		\partial_{t_{-1}'} w=\frac{\alpha_{-1}}{r_0}\left(1+r_0we^{r_0\partial_x^{-1}w}\partial_x^{-1}  e^{-r_0\partial_x^{-1}w}\right),
	\end{equation}
	\begin{equation}
		\partial_{t_{-2}'} w=\frac{\alpha_{-2}}{r_0}e^{r_0\partial_x^{-1}w}\left(\partial_x^{-1} e^{-r_0\partial_x^{-1}w}+r_0w\partial_x^{-2}  e^{-r_0\partial_x^{-1}w}\right).
	\end{equation}
	
	Eqn.  (\ref{eq11})  can be  re-written in a  local form as,
	\begin{equation}
		\partial_{t_{-1}'}\partial_x w=\frac{\partial_x w}{w}\left(\partial_{t_{-1}'} w-\frac{\alpha_{-1}}{r_0}\right)+r_0w\partial_{t_{-1}'} w.
	\end{equation}
	
	\section{The dressing method and tau functions for CLL}
	\label{sec.dressing}
	
	In this section we employ the Dressing method \cite{babelon_dressing_1992,babelon_affine_1993, ferreira_tau-functions_1997} in order to generate systematically the soliton solutions for the entire (positive and negative flows) CLL hierarchy.  The method relies upon a particular vacuum solution which could be chosen to be zero or constant nonzero vacuum solution. The method involves the construction of vertex operators from the Heisenberg operators describing the various vacuum configurations for the two dimensional gauge potentials (\ref{vac-pos})-(\ref{vac-ppos}) or (\ref{vac-neg})-(\ref{vac-nneg}).  Their eigenvalues defines their space-time dependence. In fact we shall see that there will be two types of vertices related to eigenvalues of opposite signs.  The {\it class A} is constructed out of products of the same vertex and {\it class B} constructed out of products of opposite sign vertices
	\cite{aratyn_generalized_2025}. For the CLL hierarchy with zero
	vacuum solutions only class B allows non-trivial solutions.
	For nonzero vacuum, both cases allow non-trivial soliton
	solutions and class A leads to the Burgers solutions.
	
	\subsection{Dressing transformation}
	
	In order to employ the dressing method to generate soliton solutions we shall upgrade the affine algebra  to include central terms. This is  necessary to ensure  highest weight states.  This implies  the following modification 
	\begin{align}
		A_{x} \rightarrow A_{x}-\frac{1}{2}\left(\partial_x\nu\right) \hat{c}, \qquad A_{t_{\pm N}} \rightarrow A_{t_{\pm N}}-\frac{1}{2}\left(\partial_{t_{\pm N}}\nu\right)\hat{c},
	\end{align}
	where $\nu=\nu\left(x,t_{\pm N}\right)$ is  an extra  field that vanishes  in vacuum limit and  $\hat c$ commutes with all generators of $\hat{\mathcal{G}}$.
	
	The Lax operators for the  CLL hierarchy in  vacuum, can be written as	
	\begin{align}
		A_{t_N}^{vac}&=\Sigma^{(2N)}+b\sum_{i=1}^{N-1}\left(-r_0s_0\right)^{i}\Sigma^{(2N-2 i)}, \\
		A_{t_{-N}}^{vac}&=\Upsilon^{(-2N)}+b\sum_{i=1}^{N-1}\left(-r_0s_0\right)^{-i}\Upsilon^{(-2N+2 i)}.
	\end{align}
	where $b=0$ for zero vacuum and $b=1$ for the constant nonzero vacuum.
	We consider the g-RHB decomposition proposed in \cite{aratyn_generalized_2025}
	\begin{equation}
		\label{g-RHB}
		\Theta_-^{-1}(t)\, \Theta_+(t) = \Psi_a(t) \, g \, \Psi_a^{-1}(t),
	\end{equation}
	where, $\Psi$ is the generalized Baker-Akhiezer function \eqref{B-A} with $a=2$,
	\begin{equation}
		\Psi=\exp \left [-\sum_{N=1}^{\infty } \left ( A_{t_N}^{vac} t_N+A_{t_{-N}}^{vac} t_{-N} \right ) \right ],
	\end{equation}
	and $g=e^{Y}$, where $Y$ is an arbitrary constant element of $\hat{\mathcal G}$. The left-hand side of  (\ref{g-RHB}) can be in general written as {\footnote{ In general  we may consider an asymmetric splitting of the zero grade component $\theta^ {(0)}$, i.e., $\Theta_+=e^{(1-c)\theta^{(0)}}\prod_{i=1}^{\infty} e^{\theta^{(i)}}, 
			\qquad 
			\Theta_-=e^{-c\theta^{(0)}}\prod_{i=1}^{\infty} e^{-\theta^{(-i)}}$. Here we consider $c=1/2.$}}
	\begin{align*}
		\begin{split}
			\Theta_+=e^{\frac{1}{2}\theta^{(0)}}\prod_{i=1}^{\infty} e^{\theta^{(i)}},
		\end{split} 
		\begin{split}
			\Theta_-=e^{-\frac{1}{2}\theta^{(0)}}\prod_{i=1}^{\infty} e^{-\theta^{(-i)}},
		\end{split}
	\end{align*}
	where $\theta^{(j)} \in \hat{\mathcal{G}}_j$.  In particular,  \footnote{The term $\delta_{k,0}$ denotes the Kronecker delta.}
	\begin{align*}
		\begin{split}
			\theta^{(2k)}=\varphi_{2k} h^{(k)}+\delta_{k,0} \nu \hat{c},
		\end{split} 
		\begin{split}
			\theta^{(2k+1)}=\chi_{2k+1} E_{\alpha}^{(k)}+\psi_{2k+1} E_{-\alpha}^{(k+1)},
		\end{split}
		\qquad 
		k \in \mathbb{Z}
	\end{align*}
	The coefficients $\nu$, $\varphi_{2k}$, $\chi_{2k+1}$, e $\psi_{2k+1}$, known 
	as auxiliary fields are functionals of  $x$ and $t_{\pm N}$.

	The dressing operators $\Theta_+$ and  $\Theta_-$ gauge transform the  Lax operators $A^{vac}_x=- \left(\pa_{x} \Psi\right) \Psi^ {-1}$ and $A^{vac}_{t_{\pm N}}- \left(\pa_{t_{\pm N}} \Psi\right)\Psi^ {-1}$ into  its non-trivial configuration $A_x$ and $A_{t_{\pm N}}$, i.e.,
	\begin{subequations}
		\begin{align}
			A_{x}&=\Theta_\pm A_{x}^{vac}\Theta_\pm^{-1}-\left( \partial_x \Theta_\pm \right)\Theta_\pm^{-1}=-\left[\partial_x\left(\Theta_\pm \Psi\right)\right]\left( \Theta_\pm \Psi\right)^{-1},
			\label{T_Ax}
			\\[2.5mm]
			A_{t_{\pm N}}&=\Theta_\pm A_{t_{\pm N}}^{vac}\Theta_\pm^{-1}-\left( \partial_{t_{\pm N}} \Theta_\pm \right)\Theta_\pm^{-1}=-\left[\partial_{t_\pm N}\left(\Theta_\pm \Psi\right)\right]\left(\Theta_\pm \Psi\right)^{-1}.
			\label{T_At}
		\end{align}
	\end{subequations}
	
	Solving eqns. (\ref{T_Ax}) and (\ref{T_At})  recursively we determine the auxiliary fields $ \theta^ {(\pm i)}$ in terms of the physical fields $r(x, t_{\pm N})$ and  $s(x, t_{\pm N})$ defined in \eqref{Ax1}.
	
	Decomposing  (\ref{T_Ax}) using  $\Theta_+$ we obtain from zero grade projection,
	\begin{equation} \label{phi0}
		\partial_x \varphi_0=rs-br_0s_0.
	\end{equation}
	Grade one projection yields,
	\begin{align*}
		\begin{split}
			\partial_x \chi_1-br_0s_0\chi_1=- re^{-\varphi_0}+br_0,
		\end{split} 
		\begin{split}
			\partial_x \psi_1+br_0s_0\psi_1=- se^{\varphi_0}+bs_0.
		\end{split}
	\end{align*}
	and so on in order to determine  higher order coefficients in $\Theta_+$.
	
	For transformation  $\Theta_-$, we find	from  (\ref{T_Ax}), 
	\begin{align}
		\label{campos}
		\chi_{-1}=re^{\varphi_0}-br_0, \qquad
		\psi_{-1}=-se^{-\varphi_0}+bs_0.
	\end{align}
	together with
	\begin{equation}
		\varphi_{-2}=-\nu_x-\frac{1}{2}\left(r  e^{\varphi_0}-br_{0}\right)\left(s  e^{-\varphi_0}-bs_{0}\right)-bs_{0}\left(r  e^{\varphi_0}-br_{0}\right).
	\end{equation}
	and so on until $\Theta_-$ is determined.
	
	Conversely, eqns.  (\ref{campos}) allow determining  fields $r$ and $s$  in terms  of $\varphi_0$, $\chi_{-1}$ e $\psi_{-1}$, 
	\begin{align}\label{r_s_campos_aux}
		r=\left(br_{0}+\chi_{-1}\right)e^{-\varphi_0}, 
		\qquad
		s=\left(bs_{0}-\psi_{-1}\right)e^{\varphi_0}.
	\end{align}
	
	\subsection{Tau functions}
	
	In order to determine soliton solutions  within the dressing method we introduce the $\tau-$ functions defined as
	\begin{align}\label{tau}
		\tau_{kl}\equiv \bra{\lambda_{k}}\Theta_-^{-1}\Theta_+\ket{\lambda_{l}}=\bra{\lambda_{k}}\Psi g \Psi^{-1}\ket{\lambda_{l}}, \qquad k,l=0,1,2,3,
	\end{align}
	where the  states $\ket{\lambda_{k}}$ and  $\ket{\lambda_{l}}$
	are defined as 
	\begin{align*}
		\ket{\lambda_0}=\ket{\mu_0}, \quad \quad
		\ket{\lambda_1}=\ket{\mu_1},  \quad \quad
		\ket{\lambda_2}=E^{(-1)}_{\alpha}\ket{\mu_0},  \quad \quad
		\ket{\lambda_3}=E^{(0)}_{-\alpha}\ket{\mu_1},
	\end{align*}
	with $\ket{\mu_0}$ and $\ket{\mu_1}$ being the highest
	weight states of $\hat{A}_1$.
	
	From the left-hand-side of (\ref{tau}), we can define,
	\begin{align*}
		\tau_{00}&=\bra{\mu_0}\cdots e^{\theta^{(-1)}}e^{\theta^{(0)}}e^{\theta^{(1)}} \cdots \ket{\mu_0}=\bra{\mu_0}e^{\nu\hat{c}}\ket{\mu_0}=e^{\nu}, 
		\\
		\tau_{11}&=\bra{\mu_1}\cdots e^{\theta^{(-1)}}e^{\theta^{(0)}}e^{\theta^{(1)}} \cdots \ket{\mu_1}=\bra{\mu_1}e^{\varphi_0 h^{(0)}+\nu\hat{c}}\ket{\mu_1}=e^{\varphi_0+\nu}, \\
		\tau_{20}&=\bra{\mu_0}E^{1}_{-\alpha} \cdots e^{\theta^{(-1)}}e^{\theta^{(0)}}e^{\theta^{(1)}} \cdots  \ket{\mu_0}=-e^{\nu}\bra{\mu_0}\left [ \theta^{(-1)},E^{(1)}_{-\alpha} \right  ]\ket{\mu_0}=\chi_{-1}e^{\nu}, 
		\\
		\tau_{31}&=\bra{\mu_0}E^{0}_{\alpha} \cdots e^{\theta^{(-1)}}e^{\theta^{(0)}} \cdots  \ket{\mu_1}=-e^{\varphi_0+\nu} \bra{\mu_1}\left [ \theta^{(-1)},E^{(0)}_{\alpha} \right ]\ket{\mu_1}=\psi_{-1}e^{\varphi_0+\nu}, 
	\end{align*}
	
	The following relations follow straightforwardly,
	\begin{align} \label{tau.fields}
		e^{\nu}=\tau_{00}, \qquad e^{\varphi_0+\nu}=\tau_{11}, \qquad \psi_{-1}=\frac{\tau_{31}}{\tau_{11}} \qquad \chi_{-1}=\frac{\tau_{20}}{\tau_{00}}.
	\end{align}
	Substituting these values into
	(\ref{r_s_campos_aux}), we find fields $r$ and  $s$ in
	terms of the $\tau$-functions  $\tau_{00}$,
	$\tau_{11}$, $\tau_{20}$, and $\tau_{31}$,
	\begin{align}\label{r_s_tau}
		r=\frac{b r_{0}\tau_{00}+\tau_{20}}{\tau_{11}}, \qquad s=\frac{b s_{0}\tau_{11}-\tau_{31}}{\tau_{00}}.
	\end{align}

	\subsection{Vertex operators}
	
	An important ingredient in constructing and
	classifying solutions are the vertex operators. These
	are the eigenstates of the Heisenberg sub-algebras whose
	eigenvalues lead to the space-time dependence of the
	solitons for the entire hierarchy, i.e., 
	\begin{align}\label{cmV}
		\comm{V_i^{\pm}}{A_{x}^{vac}}=\pm \kappa_x V_i^{\pm}, \qquad \comm{V_i^{\pm}}{A_{t_{\pm N}}^{vac}}=\pm \omega_{\pm N} V_i^{\pm}. 
	\end{align}
	It can be checked that {\footnote{ Notice that for $b=1$  these correspond to deformed  vertex operators depending upon parameters $r_0$ and $s_0$.}}
	\begin{equation}\label{V+}
		V_i^+\equiv V^+\left( k_i\right)=-b  r_{0}\hat{c}+\sum_{j=-\infty }^{\infty }\left(b r_{0} k_i^{-j}h^{(j)}+b r_{0}^2k_i^{-j}E_{\alpha}^{(j-1)}-k_i^{-j+1}E_{-\alpha}^{(j)}\right),
	\end{equation}
	\begin{equation}\label{V-}
		V_i^-\equiv V^-\left( k_i\right)=\sum_{j=-\infty }^{\infty }\left ( b  s_{0} k_i^{-j}h^{(j)}-k_i^{-j+1}E_{\alpha}^{(j-1)}+b s_{0}^2k_i^{-j}E_{-\alpha}^{(j)}\right ),
	\end{equation}
	where  $k_i$ is a complex parameter, with $i \in \mathbb{Z} $,
	that satisfy (\ref{cmV})  with 
	\begin{align} \label{eigenvalues}
		\kappa_x=k_i+ b r_{0}s_{0}, \qquad \omega_N=k_i^N- b \left(- r_{0}s_{0}\right)^N, \qquad \omega_{-N}=\left( 1-2b\right)k_i^{-N}+ b\left(- r_{0}s_{0}\right)^{-N}.
	\end{align}
	It therefore follows that
	\begin{equation}
		\Psi V_i^{\pm}\Psi^{-1}=V_i^{\pm}+\left[  V_i^{\pm},  A_{x}^{vac} x+ A_{t_{\pm N}}^{vac} t_{\pm N}\right]+\frac{1}{2}\left[ \left[  V_i^{\pm}, A_{x}^{vac} x+ A_{t_{\pm N}}^{vac} t_{\pm N}\right],  A_{x}^{vac} x+ A_{t_{\pm N}}^{vac} t_{\pm N} \right]+\cdots,
	\end{equation}
	and
	\begin{align}\label{Vk}
		\Psi V_i^{\pm}\Psi^{-1}=\rho_i V_i^{\pm}, \qquad  \rho_i=\rho_i(x,t_{\pm N})=e^{\kappa_x x + \omega_{\pm N} t_{\pm N}}.
	\end{align}
	Using the identity $\Psi^{-1}\Psi=1$, it follows that
	\begin{equation}
		\Psi \left (V_i^{\pm}\right )^n\Psi^{-1}=\left (\Psi V_i^{\pm}\Psi^{-1}\right )^n
	\end{equation}
	and hence,
	\begin{equation}
		\Psi e^{V_i^{\pm}} \Psi^{-1}=\exp \left( \rho_i V_i^{\pm} \right).
	\end{equation}
	The $\tau$ functions (\ref{tau}) can be  exactly evaluated by choosing  $g=\prod\limits_{i=1}^{n}e^{V_i^{\pm}}$.

	\subsection{Class A and solitons for Burgers hierarchy}
	Assuming $g=\prod\limits_{i=1}^{n}e^{V_i^{\pm}}$,  we obtain  a class of solutions involving products  of  single vertices, either $V_i^+$ or $V_i^-$,
	\begin{equation}
		\label{CA}
		\tau_{kl}=\bra{\lambda_{k}} \prod_{i=1}^{n}\left( 1+\rho_iV_i^{+}+\rho_i^2\left(V_i^+\right)^2+\cdots \right) \ket{\lambda_{l}}.
	\end{equation}
	Evaluating the  $\tau-$functions  $\tau_{00}$, $\tau_{11}$, $\tau_{20}$, and $\tau_{31}$, 
	\begin{align}\label{solA}
		\tau_{00}=1-b r_0 \sum_{i=1}^{n}\rho_i, \qquad
		\tau_{11}=1, \qquad
		\tau_{20}=b r_0^2 \sum_{i=1}^{n}\rho_i,\qquad
		\tau_{31}=-\sum_{i=1}^{n}k_i\rho_i.
	\end{align}
	Substituting in  (\ref{r_s_tau}) we find for general values of n,
	\begin{align} 
		r= b r_0, \qquad s=\frac{b s_{0}+\displaystyle\sum_{i=1}^{n}k_i\rho_i}{1-b r_{0}\displaystyle\sum_{i=1}^{n}\rho_i}.
	\end{align}
	For the particular case where  $b=0$, we find  the trivial wave solution  for the associated heat equation  for field $\phi$ (see table \ref{tab:reducoesCLL}),
	\begin{align}
		r\to 0, \qquad s\to\phi=\sum_{i=1}^{n}k_i\exp\left\{k_i\,x+\left(k_i\right)^{\pm N}t_{\pm N}\right\},
	\end{align}
	For  $b=1$
	\begin{align} \label{w}
		r\to r_0, \qquad s\to w=\frac{ s_{0}+\displaystyle\sum_{i=1}^{n}k_i\exp\left\{\left(k_i+r_0s_0\right)x\pm\left[\left(k_i\right)^{\pm N}-\left(-r_0s_0\right)^{\pm N}\right]t_{\pm N}\right\}}{1- r_{0}\displaystyle\sum_{i=1}^{n}\exp\left\{\left(k_i+r_0s_0\right)x\pm\left[\left(k_i\right)^{\pm N}-\left(-r_0s_0\right)^{\pm N}\right]t_{\pm N}\right\}},
	\end{align}
	we find $w$  in (\ref{w}) to solve the Burgers  hierarchy. 
	
	Re-writing  the $n$-solitons solution as 
	\begin{align}\label{swn}
		w=-\left( r_0\right)^{-1}\frac{ \partial_x \Phi}{\Phi}, \; \Phi=\exp\left\{-r_0s_0\,x\pm\alpha_{\pm N}\left(-r_0s_0\right)^{\pm N}t_{\pm N}'\right\}-r_0\displaystyle\sum_{i=1}^{n}\exp\left\{k_ix\pm\alpha_{\pm N}\left(k_i\right)^{\pm N}t_{\pm N}'\right\},
	\end{align}
	we can express $w$ in terms of  variable $\Phi=\Phi\left(x, t_{\pm N}\right)$
	satisfying 
	\begin{equation}
		\partial_{t_N}\Phi=\alpha_{N}\partial_x^N \Phi, \qquad \partial_{t_{-N}}\Phi=-\alpha_{-N}\partial_x^{-N} \Phi,
	\end{equation}
	via the Cole-Hopf transformation.
	We should point out that the Cole-Hopf transformation \cite{hopf_partial_1950, cole_quasi-linear_1951}, was employed to all positive flows of the Burgers hierarchy by 
	Kudryashov \cite{kudryashov_exact_2009}. Later in \cite{aratyn_generalized_2025}  it  was extended to all negative  sub-hierarchy. In fact this was shown to be  realized as a gauge transformation of Miura type between CLL and AKNS hierarchies.
	
	Exchanging $V_i^+$ for  $V_i^-$ in (\ref{CA}) we find a similar result after exchanging $r\to s$ and $\rho_i \to - \rho_i^{-1}$.

	\subsection{Class B and solitons for CLL hierarchy}
	Let us now consider products of mixed vertices, $g=\prod\limits_{i=1}^{n}e^{V_i^{+}}e^{V_{i+1}^{-}}$ such that, 
	\begin{equation}
		\label{CB}
		\tau_{kl}=\bra{\lambda_{k}} \prod_{i=1}^{n}\left( 1+\rho_i V_i^{+}+\rho_{i+1}^{-1}V_{i+1}^{-}+\rho_i\rho_{i+1}^{-1}V_i^{+}V_{i+1}^-+\cdots \right)\ket{\lambda_{l}}.
	\end{equation}
	Evaluating  $\tau_{00}$, $\tau_{11}$, $\tau_{20}$, and  $\tau_{31}$  we find for $n=1$, 
	\begin{subequations}
		\begin{align}\label{solB}
			\tau_{00}&=1-br_0\rho_1+\frac{k_2\left(k_1+b  r_{0}s_{0}\right)^2}{\left(k_2-k_1\right)^2}\rho_1 \rho_2^{-1},\\
			\tau_{11}&=1+bs_0\rho_2^{-1}+\frac{k_1\left(k_2+b  r_{0}s_{0}\right)^2}{\left(k_2-k_1\right)^2}\rho_1 \rho_2^{-1}, \\
			\tau_{20}&=b r_0^2 \rho_1-k_2\rho_2^{-1}+\frac{ b r_{0} k_2\left(k _1 +k _2+2 b r_{0}s_{0}\right)}{k _2-k _1}\rho_1 \rho_2^{-1}, \\
			\tau_{31}&=-k_1 \rho_1+bs_0^2\rho_2^{-1}+\frac{ b s_{0} k _1 \left(k _1+k _2+2 b r_{0}s_{0}\right)}{k _2-k _1}\rho_1 \rho_2^{-1}.
		\end{align}
	\end{subequations}
	
	Substituting these relations in  (\ref{r_s_tau}),  
	we obtain the 2-soliton  solution for the CLL hierarchy
	\begin{align}
		r= \frac{b r_{0}-k _2\rho_2^{-1} +  \frac{b r_{0} k _2 \left(k _2+br_{0} s_{0}\right)^2}{\left(k _2-k _1\right)^2}\rho_1 \rho_2^{-1}}{1+bs_{0} \rho_2^{-1}+  \frac{k _1\left(k _2+br_{0} s_{0}\right)^2}{\left(k _2-k _1\right)^2} \rho_1 \rho_2^{-1}}, \qquad s=\frac{bs_{0}+k _1 \rho_1+ \frac{b s_{0}k _1\left(k _1+b r_{0} s_{0}\right)^2}{\left(k _2-k _1\right)^2} \rho_1 \rho_2^{-1}}{1-b r_{0} \rho_1+ \frac{k _2\left(k _1+b r_{0} s_{0}\right)^2}{\left(k _2-k _1\right)^2} \rho_1 \rho_2^{-1}}.
	\end{align}
	for $b=0$ or $b=1$.
	
	Conversely, exchanging  $e^{V_1^+}e^{V_{2}^-}$ with
	$e^{V_1^-}e^{V_{2}^+}$, we find  another  pair of solutions, 
	\begin{align}
		r = \frac{b r_{0}-k_1 \rho_1^{-1}+ \frac{b r_{0}k_1\left(k _1+b r_{0} s_{0}\right)^2}{\left(k _2-k _1\right)^2}\rho_1^{-1}\rho_2}{1+bs_0\rho_1^{-1}+\frac{k_2\left(k_1+b  r_{0}s_{0}\right)^2}{\left(k_2-k_1\right)^2}\rho_1^{-1}\rho_2}, \qquad s=\frac{b s_{0}+k_2\rho_2+\frac{b s_{0}k _2\left(k _2+b r_{0} s_{0}\right)^2}{\left(k _2-k _1\right)^2}\rho_1^{-1}\rho_2}{1-br_0 \rho_2+\frac{k_1\left(k_2+b  r_{0}s_{0}\right)^2}{\left(k_2-k_1\right)^2}\rho_1^{-1}\rho_2}.
	\end{align}

	\section{Gauge-Bäcklund transformation}
	\label{sec.backlund}
	
	Bäcklund transformations play an important role in the construction and characterization of solutions in integrable systems. These transformations may be obtained through several formulations and techniques \cite{rogers_backlund_1982}. In this work, we formulate the Bäcklund transformations as gauge transformations that preserve the zero curvature as this property ensures that the resulting relations \textit{extend to all flows}. 
	
	This universality arises from the fact that all flows of a given integrable hierarchy share the same underlying algebraic structure and possess a common Lax pair, whose spatial component we denote by $A_x$. The special case in which it connects two configurations within the same equation of motion is referred to as auto-Bäcklund transformations and we shall explore it for the CLL hierarchy case in the present section. Within the algebraic framework, the Bäcklund transformation can be represented by a \emph{gauge} transformation, since the zero-curvature condition is gauge invariant and therefore flows equations are unchanged.
	
	Consider then the gauge-transformed Lax pair given by:
	\begin{align}\label{gauge-transformation-bäcklund}
		A_\mu (\psi)=  U\left(\phi, \psi, \lambda \right) A_\mu (\phi)  U^{-1}\left(\phi, \psi,\lambda \right) +  U \left(\phi, \psi, \lambda \right)\pa_\mu U^{-1}\left(\phi, \psi, \lambda \right)
		\qquad
		(\mu = x \; \text{or} \; t_{\pm N}).
	\end{align}
	$A_\mu (\phi)$ and $A_\mu (\psi)$ are Lax pairs in different field configurations and $U$ is a group element (expanded in terms of algebra elements), that depend on the field configurations and the spectral parameter $\lambda$.
	
	In a series of works, we developed an approach that uses the affine structure of the algebra to propose different graded ansatzes for the Bäcklund transformation. In \cite{de_carvalho_ferreira_gauge_2021, de_carvalho_ferreira_generalized_2021} we have shown that different graded ansatz are related to Type I and Type II Bäcklund transformations for the sinh-Gordon hierarchy \cite{corrigan_jump-defects_2006, corrigan_new_2009} and generalized this result to $A_r$-mKdV hierarchy. More recently, we have extended this approach to the negative sector of both mKdV \cite{adans_negative_2023}. 
	
	Let us denote the different CLL configurations as follows:
	\begin{align}
		A_\mu (\phi) \equiv A^{\text{CLL}}_\mu (r_1, s_1)
		\qquad
		\text{and}
		\qquad
		A_\mu (\psi) \equiv A^{\text{CLL}}_\mu (r_2, s_2).
	\end{align}
	such that \eqref{gauge-transformation-bäcklund} become
	\begin{align}
		A^{\text{CLL}}_\mu (r_2, s_2)\, U
		- U\, A^{\text{CLL}}_\mu (r_1, s_1)
		+ \partial_\mu U = 0,
		\label{gauge-transformation-bäcklund.2}
		\qquad
		\text{with} \qquad
		\mu = x \;\text{or}\; t_{\pm N} ,
	\end{align}
	Next, we propose an $2 \times 2$ matrix ansatz  in order to implement the gauge–Bäcklund transformation by using the graded structure present in the  $\hat{sl}(2)$  affine algebra, as in \cite{de_carvalho_ferreira_gauge_2021}. To accomplish this, we consider the following $2 \times 2$ graded matrices
	\begin{align}
		U^{(2n)} =
		\begin{pmatrix}
			\la ^{n} a_{1,1}^{(2n)} & &0
			\\
			\\
			0 & & \la ^{n} a_{2,2}^{(2n)} 
		\end{pmatrix}, \qquad 	U^{(2n+1)} =
		\begin{pmatrix}
			0 & &\la ^{n} a_{1,2}^{(2n+1)}
			\\
			\\
			\la ^{n+1} a_{1,2}^{(2n+1)} & & 0
		\end{pmatrix}
	\end{align}
	where $\la$ is the, previously introduced, spectral parameter, $u_{i,j}$ are functional of the fields $r_i, s_i$ and the upper index indicates the grade of the matrix. Then, for each  Bäcklund transformation, we consider a different expansion given by:
	\begin{center}
		\begin{tabular}{lc}
			\toprule
			\textbf{Ansatz} & \textbf{Bäcklund Transformation}
			\\
			\midrule
			$U_0=U^{(2n)}$ &	$\mathrm{B}_{\text{0}}$
			\\
			\midrule
			$U_{I}=U^{(2n)}+U^{(2n+1)}$ &	$\mathrm{B}_{\text{I}}$
			\\
			\midrule
			$U_{II}=U^{(2n)}+U^{(2n+1)}+U^{(2n+2)}$ &	$\mathrm{B}_{\text{II}}$
			\\
			\bottomrule
		\end{tabular}
	\end{center}
	to enable us to solve \eqref{gauge-transformation-bäcklund.2} for each $U_i$, determining both $a_{ij}$ and the Bäcklund transformation. In the following section, we present this procedure for ansatz II.
	We shall see, therefore, that the most relevant information in the gauge ansatz is the sum of successive graded subspaces, since different sums lead to different Bäcklund transformations.

	\subsection{Determining the transformation}
	
	We now propose the ansatz II for the gauge–Bäcklund transformation, which is the most general one and covers the previous ansatz as we take appropriated limits. For particular choice $n=-1$, it takes the form
	\begin{align} 
		U_{\text{II}} =U^{(0)}+U^{(-1)}+U^{(-2)}=
		\begin{pmatrix}
			a_{1,1} + \dfrac{1}{\la}\, b_{1,1} & \dfrac{1}{\la}\, a_{1,2} \\
			a_{2,1} & a_{2,2} + \dfrac{1}{\la}\, b_{2,2}
		\end{pmatrix},
	\end{align}
	where we introduced $a_{ij}^{(0)}=a_{i,j}$, $a_{ij}^{(-1)}=a_{i,j}$ and $a_{ij}^{(-2)}=b_{i,j}$ to simplify the notation. Substituting this ansatz into \eqref{gauge-transformation-bäcklund.2} yields the following system of equations
	\begin{subequations}
		\begin{align}
			- a_{1,1} r_1 + a_{2,2} r_2 + a_{1,2} &= 0,
			\label{sist.eq.ansatz.III.1}
			\\[1.5mm]
			a_{1,1} s_2 - a_{2,2} s_1 - a_{2,1} &= 0,
			\label{sist.eq.ansatz.III.2}
			\\[1.5mm]
			\partial_x b_{1,1} + \tfrac{1}{2} b_{1,1}\, \partial_x (J_1 - J_2) &= 0,
			\label{sist.eq.ansatz.III.3}
			\\[1.5mm]
			\partial_x b_{2,2} - \tfrac{1}{2} b_{2,2}\, \partial_x (J_1 - J_2) &= 0,
			\label{sist.eq.ansatz.III.4}
			\\[1.5mm]
			\partial_x a_{1,1} + \tfrac{1}{2} a_{1,1}\, \partial_x (J_1 - J_2)
			+ a_{2,1} r_2 - a_{1,2} s_1 &= 0,
			\label{sist.eq.ansatz.III.5}
			\\[1.5mm]
			\partial_x a_{2,2} - \tfrac{1}{2} a_{2,2}\, \partial_x (J_1 - J_2)
			- a_{2,1} r_1 + a_{1,2} s_2 &= 0,
			\label{sist.eq.ansatz.III.6}
			\\[1.5mm]
			\partial_x a_{1,2} - \tfrac{1}{2} a_{1,2}\, \partial_x (J_1 + J_2)
			- b_{1,1} r_1 + b_{2,2} r_2 &= 0,
			\label{sist.eq.ansatz.III.7}
			\\[1.5mm]
			\partial_x a_{2,1} + \tfrac{1}{2} a_{2,1}\, \partial_x (J_1 + J_2)
			- b_{2,2} s_1 + a_{1,1} s_2 &= 0.
			\label{sist.eq.ansatz.III.8}
		\end{align}
		\label{sist.eq.ansatz.III.sol.1}
	\end{subequations}
	where $\pa_x J_i = r_i s_i$ with $i=1,2$.	By direct integration of equations \eqref{sist.eq.ansatz.III.3}–\eqref{sist.eq.ansatz.III.4} we obtain
	\begin{align}
		b_{1,1} = \gamma_1\, e^{-\tfrac{1}{2}(J_1 - J_2)}
		\qquad
		\text{and}
		\qquad
		b_{2,2} = \gamma_2\, e^{\tfrac{1}{2}(J_1 - J_2)}.
	\end{align}
	From \eqref{sist.eq.ansatz.III.1}–\eqref{sist.eq.ansatz.III.2} we can isolate
	\begin{align}
		a_{1,2} = a_{1,1} r_1 - a_{2,2} r_2,
		\qquad
		a_{2,1} = a_{1,1} s_2 - a_{2,2} s_1,
		\label{entries.a12.a21.ansatz.III}
	\end{align}
	and after substituting \eqref{entries.a12.a21.ansatz.III} into \eqref{sist.eq.ansatz.III.5} and \eqref{sist.eq.ansatz.III.6}, we obtain
	\begin{subequations}
		\begin{align}
			\partial_x a_{1,1} - \tfrac{1}{2} \partial_x (J_1 - J_2) &= 0,
			\qquad\Rightarrow\qquad
			a_{1,1} = \alpha_1\, e^{\tfrac{1}{2}(J_1 - J_2)},
			\\[2.5mm]
			\partial_x a_{2,2} + \tfrac{1}{2} \partial_x (J_1 - J_2) &= 0,
			\qquad\Rightarrow\qquad
			a_{2,2} = \alpha_2\, e^{-\tfrac{1}{2}(J_1 - J_2)}.
		\end{align}
		\label{entries.a11.a22.ansatz.III}
	\end{subequations}
	Hence the functions \(a_{1,2}\) and \(a_{2,1}\) become
	\begin{subequations}
		\begin{align}
			a_{1,2} &= \alpha_1 e^{\tfrac{1}{2}(J_1 - J_2)} r_1
			- \alpha_2 e^{-\tfrac{1}{2}(J_1 - J_2)} r_2,
			\\[2.5mm]
			a_{2,1} &= \alpha_1 e^{\tfrac{1}{2}(J_1 - J_2)} s_2
			- \alpha_2 e^{-\tfrac{1}{2}(J_1 - J_2)} s_1.
		\end{align}
	\end{subequations}
	Having determined all entries of the matrix associated with ansatz II, the resulting gauge–Bäcklund transformation is takes the form
	\begin{align}
		U_{\text{II}} = \label{gauge.baklund.3}
		\begin{pmatrix}
			\alpha _1 e^{\frac{1}{2}\left( J_1- J_2\right)}+\frac{\gamma _1 e^{\frac{1}{2}\left(J_2-J_1\right)}}{\lambda } & &\frac{e^{-\frac{1}{2} \left(J_1+J_2\right)} \left(\alpha _1 e^{J_1} r_1-\alpha _2 e^{J_2} r_2\right)}{\lambda }
			\\
			\\
			e^{-\frac{1}{2} \left(J_1+J_2\right)} \left(\alpha _1 e^{J_1} s_2-\alpha _2 e^{J_2} s_1\right) & & \alpha _2 e^{\frac{1}{2}\left( J_2- J_1\right)}+\frac{\gamma _2 e^{\frac{1}{2}\left(J_1-J_2\right)}}{\lambda }
		\end{pmatrix},
	\end{align}
	which satisfies \eqref{gauge-transformation-bäcklund.2} provided that the following differential relations hold:
	\begin{subequations}\label{bäcklund.x.type.III.cll.determining}
		\begin{align}
			\alpha_2\, \partial_x \!\left(r_2 e^{-J_1}\right)
			+ \gamma_1 e^{-J_1} r_1
			&=
			\alpha_1\, \partial_x \!\left(r_1 e^{-J_2}\right)
			+ \gamma_2 e^{-J_2} r_2,
			\\[2mm]
			\alpha_1\, \partial_x \!\left(s_2 e^{J_1}\right)
			- \gamma_2 e^{J_1} s_1
			&=
			\alpha_2\, \partial_x \!\left(s_1 e^{J_2}\right)
			- \gamma_1 e^{J_2} s_2.
		\end{align}
	\end{subequations}
	These relations constitute the type II Bäcklund transformation or simply denoted as $B_{\text{II}}$. They contain spatial derivatives as commonly occurs in similar transformations for other models. The appropriate limits reduce the $B_{\text{II}}$ to the simplest cases, type 0 ($B_{\text{0}}$) and type I ($B_{\text{I}}$).
	
	\subsection{Reductions}
	
	\noindent 
	The equations \eqref{bäcklund.x.type.III.cll.determining} possesses two pairs of Bäcklund parameters, $\left(\al_1,\al_2\right)$ and $\left(\gamma_1,\gamma_2\right)$. 
	We now analyze how Bäcklund transformations reduces under certain limits for these parameters.
	
	\paragraph{I}
	In the limit $\alpha_i \to 0$, the transformations reduce to the simplest case, namely \(B_{\text{0}}\), such that \eqref{bäcklund.x.type.III.cll.determining} becomes
	\begin{subequations}
		\begin{align}
			\gamma_1 e^{-J_1} r_1 &= \gamma_2 e^{-J_2} r_2,
			\qquad\Rightarrow\qquad
			r_2 = \frac{\gamma_1}{\gamma_2} e^{-J_1+J_2} r_1,
			\\[2mm]
			\gamma_2 e^{J_1} s_1 &= \gamma_1 e^{J_2} s_2,
			\;\qquad\Rightarrow\qquad
			s_2 = \frac{\gamma_2}{\gamma_1} e^{J_1-J_2} s_1.
		\end{align}
	\end{subequations}
	Hence,
	\begin{align}
		r_2 s_2 = r_1 s_1\quad\Rightarrow\quad \pa_xJ_1= \pa_xJ_2\quad\Rightarrow\quad J_2 = J_1 + \delta ,
	\end{align}
	where $\delta$  is a constant. 
	Thus the Bäcklund transformations on the fields configurations may be written as
	\begin{align}
		r_2 = \frac{\gamma_1}{\gamma_2} e^{\delta} r_1,
		\qquad
		s_2 = \frac{\gamma_2}{\gamma_1} e^{-\delta} s_1,
	\end{align}
	that can be interpreted as a trivial scaling transformation. Indeed, previous work has shown that a zero  order expansion always leads to this trivial transformation. Accordingly, the corresponding gauge–Bäcklund transformation \eqref{gauge.baklund.3} reduces to
	\begin{align}\label{gauge.baklund.1}
		U_{II}
		\quad
		\underset{\alpha_i \to 0}{\longrightarrow}
		\quad
		U_{\text{0}} =
		\begin{pmatrix}
			\frac{\gamma_1}{\lambda}\, e^{\delta/2} & 0 \\
			0 & \frac{\gamma_2}{\lambda} \, e^{-\delta/2}
		\end{pmatrix}.
	\end{align}
	\paragraph{II} On the other hand, the limit $\gamma_i \to 0$ leads to \(B_{\text{I}}\), and \eqref{bäcklund.x.type.III.cll.determining} becomes
	\begin{subequations}
		\begin{align}
			\alpha_2\, \partial_x \!\left(r_2 e^{-J_1}\right)
			&=
			\alpha_1\, \partial_x \!\left(r_1 e^{-J_2}\right),
			\qquad\Rightarrow\qquad
			\alpha_2\, r_2 e^{-J_1}
			=
			\alpha_1\, r_1 e^{-J_2} + \beta_1
			\\[2mm]
			\alpha_1\, \partial_x \!\left(s_2 e^{J_1}\right)
			&=
			\alpha_2\, \partial_x \!\left(s_1 e^{J_2}\right),
			\;\; \qquad\Rightarrow\qquad
			\alpha_1\, s_2 e^{J_1}
			=
			\alpha_2\, s_1 e^{J_2} + \beta_2,
		\end{align}
	\end{subequations}
	where $\beta_1$, $\beta_2$ are constants of integration (Bäcklund parameters). The Type I Bäcklund transformation indeed gives a non-trivial transformation. It cannot be reduced to a scaling transformation as in \ref{bäcklund.x.type.I.cll} and does not depend on derivatives of fields as one usually expects for such kind of expansion. The corresponding gauge–Bäcklund transformation \eqref{gauge.baklund.3} turns out to be:
	\begin{align}\label{gauge.baklund.2}
		U_{II}
		\quad
		\underset{\gamma_i \to 0}{\longrightarrow}
		\quad
		U_{\text{I}} =
		\begin{pmatrix}
			\alpha_1 \, e^{\frac{1}{2} \left(J_1-J_2\right)} & &\frac{\beta_1 \, e^{\frac{1}{2} \left(J_1+J_2\right)}}{\lambda }
			\\
			\\
			\beta_2 \, e^{-\frac{1}{2} \left(J_1+J_2\right)} & & \alpha_2 \, e^{\frac{1}{2} \left(J_2-J_1\right)} 
		\end{pmatrix},
	\end{align}
	taking the limit $\beta_i \rightarrow 0 $ we recover the type I Bäcklund transformation from type II Bäcklund transformation as we discussed before. In this way, the type II Bäcklund transformation $B_{\text{II}}$ contains the previous cases as limits.

	\subsection{Bäcklund Transformations}
	Having obtained the gauge transformations above, we summarize our results. Each gauge-Bäcklund transformation $U_i$, \eqref{gauge.baklund.1}, \eqref{gauge.baklund.2} and \eqref{gauge.baklund.3} has a different Bäcklund transformation associated to it in order to satisfy \eqref{gauge-transformation-bäcklund}. These transformations are listed below: 
	
	\begin{enumerate}[label=(\roman*)]
		\item Type 0
		\begin{subequations}\label{bäcklund.x.type.I.cll}	
			\begin{align}
				\mathrm{B}_{\text{0}} &: \quad \gamma_2 r_2 = \gamma _1 e^{\delta } r_1, 
				\\[3mm]
				\; & \;\;\;\quad  \gamma_1 s_2 =\gamma _2 e^{-\delta } s_1,
			\end{align}
		\end{subequations}
		
		\item Type I
		\begin{subequations}\label{bäcklund.x.type.II.cll}
			\begin{align}
				\mathrm{B}_{\text{I}} &: 
				\quad 
				\alpha_2\, r_2 e^{-J_1} = \alpha_1\, r_1 e^{-J_2} + \beta_1, 
				\\[2mm]
				\; & \quad
				\quad 
				\alpha_1\, s_2 e^{J_1} = \alpha_2\, s_1 e^{J_2} + \beta_2.		
			\end{align}
		\end{subequations}		
		
		\item Type II
		\begin{subequations}\label{bäcklund.x.type.III.cll}
			\begin{align}
				\mathrm{B}_{\text{II}} : \quad &\alpha_2\, \partial_x \!\left(r_2 e^{-J_1}\right)
				+ \gamma_1 e^{-J_1} r_1
				=
				\alpha_1\, \partial_x \!\left(r_1 e^{-J_2}\right)
				+ \gamma_2 e^{-J_2} r_2
				\\[2mm]
				\quad &\alpha_1\, \partial_x \!\left(s_2 e^{J_1}\right)
				- \gamma_2 e^{J_1} s_1
				=
				\alpha_2\, \partial_x \!\left(s_1 e^{J_2}\right)
				- \gamma_1 e^{J_2} s_2.
			\end{align}
		\end{subequations}
	\end{enumerate}
	Finally, using any of the $U_i$, the gauge-transformation can be applied to the temporal Lax operators:
	\begin{align}\label{gauge-transformation-bäcklund.t}
		A^{\text{CLL}}_{t_N} (r_2, s_2)\, U_i - U_i\, A^{\text{CLL}}_{t_N} (r_1, s_1) + \partial_{t_N} U_i = 0.
	\end{align}
	This equation is satisfied using only Bäcklund transformation and the equation of motion corresponding for each $U_i$ and flow $t_N$. 
	\textit{This reinforces the universality of the Bäcklund transformation within the hierarchy}.

	
	\section{Bäcklund transformations and integrable defects}
	\label{sec.defect}
	
	Integrable defects, or ``jump-defects'', can be understood as localized discontinuities that connect the fields of the model on both sides of the defect while preserving integrability. It is well established that jump-defects in integrable systems are typically related to Bäcklund transformations frozen at the defect position. This framework of Bäcklund transformations is particularly relevant, since our main interest lies in the interaction between solitons and such defects. A large body of work has investigated integrable defects in several models, including KdV, mKdV, sinh-Gordon, sine-Gordon, Tzitzéica, Boussinesq, nonlinear Schrödinger equation (NLS), among others \cite{bowcock_affine_2004, bowcock_classically_2004, corrigan_jump-defects_2006, caudrelier_systematic_2008, corrigan_new_2009, caudrelier_multisymplectic_2015, corrigan_type_2018, corrigan_adding_2023}.
	
	In our case, we assume that Bäcklund transformations introduced in the previous section and formulated as gauge transformations describe a ``jump-defect'' located at a fixed position in the CLL hierarchy. We consider a defect localized at $x=x_0$, in the sense of \cite{bowcock_affine_2004}.  Let $\left(r_1,s_1\right)$ be the field configuration before the defect, $x<x_0$ and  $\left(r_2,s_2\right)$ after the defect, $x>x_0$ for the CLL flow $t_N$.
	At the defect point, the fields satisfy the matching conditions:
	\begin{subequations}
		\begin{align}
			 \quad &\alpha_2\, \partial_x \!\left(r_2 e^{-J_1}\right)
			+ \gamma_1 e^{-J_1} r_1
			=
			\alpha_1\, \partial_x \!\left(r_1 e^{-J_2}\right)
			+ \gamma_2 e^{-J_2} r_2
			\\[2mm]
			\quad &\alpha_1\, \partial_x \!\left(s_2 e^{J_1}\right)
			- \gamma_2 e^{J_1} s_1
			=
			\alpha_2\, \partial_x \!\left(s_1 e^{J_2}\right)
			- \gamma_1 e^{J_2} s_2 , \quad x= x_0
		\end{align}
	\end{subequations}
	These equations show that soliton interactions are localized at the defect and are encoded by Bäcklund transformations.

	In order to analyze the interaction between solitons and defects, we reformulate the Bäcklund transformations in terms of tau functions, which provide a more efficient computational framework. We then present the soliton solutions before and after the defect, and investigate the conditions under which the Bäcklund transformations are satisfied.
	
	As previously stated, $B_\text{II}$ contains the remaining classes of Bäcklund transformations, so we restrict our analysis to this type, since the other cases can be recovered as reductions of this one

	\subsection{Bäcklund transformations and $\tau$-functions}
	
	The structure of the CLL fields written in terms of tau functions was previously determined via the dressing method in the previous section \ref{sec.dressing}. For instance, we recall that  \eqref{phi0}, \eqref{tau.fields} and \eqref{r_s_tau} determine
	\begin{align}
		r = \frac{br_0 \tau_{0,0} + \tau_{2,0}}{\tau_{1,1}},
		\quad
		s = \frac{bs_0 \tau_{1,1} - \tau_{3,1}}{\tau_{0,0}},
		\quad
		J = \ln\!\left(\frac{\tau_{1,1}}{\tau_{0,0}}\right) -b r_0 s_0 x,
		\label{rsJ.tau}
	\end{align}
	where $\tau_{i,j} = \tau_{i,j}(x,t_m)$ and $(r_0,s_0)$ denotes the vacuum pair. 
	
	Since the Bäcklund transformations relate a pair of solution $(r_1,s_1)$ to another pair of solution $(r_2,s_2)$,  we will  adopt the following notation for each configuration
	\begin{center}
			\begin{tabular}{lc}
				\toprule
				Solution & Tau Function \\
				\midrule
				$(r_1, s_1)$ & $(\tau_{i,j}, r_0, s_0)$ 
				\\[3mm]
				$(r_2, s_2)$ & $(\bar{\tau}_{i,j}, \bar{r}_0, \bar{s}_0)$
				\\
				\bottomrule
			\end{tabular}
	\end{center}
	so that we can reformulate the Bäcklund transformation as $\tau$-functions. Accordingly, the Bäcklund transformations originally expressed as functional of the fields $r$, $s$ and the Bäcklund parameters, are here rewritten in terms of tau functions as:
	
	\begin{subequations}\label{bäcklund.x.type.III.cll.tau}
		\begin{align}
			\mathcal{B}_{\text{II}} : \quad
			\Gamma_1& \; \left(
			\alpha_1 b r_0 \bar{r}_0 \bar{s}_0 \tau_{0,0} \tau_{1,1} \bar{\tau}_{0,0} \bar{\tau}_{1,1}
			+ \alpha_1 b \bar{r}_0 \bar{s}_0 \tau_{1,1} \tau_{2,0} \bar{\tau}_{0,0} \bar{\tau}_{1,1}
			- \alpha_1 b r_0 \tau_{1,1} \pa_x \tau_{0,0} \bar{\tau}_{0,0} \bar{\tau}_{1,1}
			\right. 
			\nonumber
			\\
			&\left. 
			+  \; \alpha_1 b r_0 \tau_{0,0} \pa_x \tau_{1,1} \bar{\tau}_{0,0} \bar{\tau}_{1,1}
			- \alpha_1 b r_0 \tau_{0,0} \tau_{1,1} \bar{\tau}_{1,1} \pa_x \bar{\tau}_{0,0}
			+ \alpha_1 b r_0 \tau_{0,0} \tau_{1,1} \bar{\tau}_{0,0} \pa_x \bar{\tau}_{1,1} 
			\right. 
			\nonumber
			\\
			&\left. 
			-  \; b \gamma_2 \bar{r}_0 \tau_{1,1}^2 \bar{\tau}_{0,0}^2
			+ \alpha_1 \tau_{2,0} \pa_x \tau_{1,1} \bar{\tau}_{0,0} \bar{\tau}_{1,1}
			- \alpha_1 \tau_{1,1} \pa_x \tau_{2,0} \bar{\tau}_{0,0} \bar{\tau}_{1,1}
			\right. 
			\nonumber
			\\
			&\left. 
			-  \; \alpha_1 \tau_{1,1} \tau_{2,0} \bar{\tau}_{1,1} \pa_x \bar{\tau}_{0,0}
			+ \alpha_1 \tau_{1,1} \tau_{2,0} \bar{\tau}_{0,0} \pa_x \bar{\tau}_{1,1}
			- \gamma_2 \tau_{1,1}^2 \bar{\tau}_{0,0} \bar{\tau}_{2,0} \right) 
			\nonumber
			\\[0.01cm]
			+\; \Gamma_2 &
			\left( \alpha_2 b \bar{r}_0 \tau_{1,1} \pa_x \tau_{0,0} \bar{\tau}_{0,0} \bar{\tau}_{1,1} - \alpha_2 b r_0 s_0 \bar{r}_0 \tau_{0,0} \tau_{1,1} \bar{\tau}_{0,0} \bar{\tau}_{1,1}
			- \alpha_2 b r_0 s_0 \tau_{0,0} \tau_{1,1} \bar{\tau}_{1,1} \bar{\tau}_{2,0}
			\right. 
			\nonumber
			\\
			&\left.
			- \;\alpha_2 b \bar{r}_0 \tau_{0,0} \pa_x \tau_{1,1} \bar{\tau}_{0,0} \bar{\tau}_{1,1}
			+ \alpha_2 b \bar{r}_0 \tau_{0,0} \tau_{1,1} \bar{\tau}_{1,1} \pa_x \bar{\tau}_{0,0}
			- \alpha_2 b \bar{r}_0 \tau_{0,0} \tau_{1,1} \bar{\tau}_{0,0} \pa_x \bar{\tau}_{1,1}
			\right. 
			\nonumber
			\\
			&\left.
			+\; b \gamma_1 r_0 \tau_{0,0}^2 \bar{\tau}_{1,1}^2
			+ \alpha_2 \tau_{1,1} \pa_x \tau_{0,0} \bar{\tau}_{1,1} \bar{\tau}_{2,0}
			- \alpha_2 \tau_{0,0} \pa_x \tau_{1,1} \bar{\tau}_{1,1} \bar{\tau}_{2,0} 
			\right. 
			\nonumber
			\\
			& \left.- \;\alpha_2 \tau_{0,0} \tau_{1,1} \bar{\tau}_{2,0} \pa_x \bar{\tau}_{1,1}
			+ \alpha_2 \tau_{0,0} \tau_{1,1} \bar{\tau}_{1,1} \pa_x \bar{\tau}_{2,0}
			+ \gamma_1 \tau_{0,0} \tau_{2,0} \bar{\tau}_{1,1}^2 \right)  = 0,
			\\[4mm]
			\quad \Gamma_1 &
			\left(
			\alpha_1 b r_0 s_0 \bar{s}_0 \tau_{0,0} \tau_{1,1} \bar{\tau}_{0,0} \bar{\tau}_{1,1}
			- \alpha_1 b r_0 s_0 \tau_{0,0} \tau_{1,1} \bar{\tau}_{0,0} \bar{\tau}_{3,1}
			- \alpha_1 b \bar{s}_0 \tau_{1,1} \pa_x \tau_{0,0} \bar{\tau}_{0,0} \bar{\tau}_{1,1}
			\right. 
			\nonumber
			\\
			&\left.
			+  \; \alpha_1 b \bar{s}_0 \tau_{0,0} \pa_x \tau_{1,1} \bar{\tau}_{0,0} \bar{\tau}_{1,1}
			- \alpha_1 b \bar{s}_0 \tau_{0,0} \tau_{1,1} \bar{\tau}_{1,1} \pa_x \bar{\tau}_{0,0}
			+ \alpha_1 b \bar{s}_0 \tau_{0,0} \tau_{1,1} \bar{\tau}_{0,0} \pa_x \bar{\tau}_{1,1} 
			\right. 
			\nonumber
			\\
			&\left.
			-  \; b \gamma_2 s_0 \tau_{1,1}^2 \bar{\tau}_{0,0}^2
			+ \alpha_1 \tau_{1,1} \pa_x \tau_{0,0} \bar{\tau}_{0,0} \bar{\tau}_{3,1}
			- \alpha_1 \tau_{0,0} \pa_x \tau_{1,1} \bar{\tau}_{0,0} \bar{\tau}_{3,1}
			\right. 
			\nonumber
			\\
			&\left.
			+  \; \alpha_1 \tau_{0,0} \tau_{1,1} \bar{\tau}_{3,1} \pa_x \bar{\tau}_{0,0}
			- \alpha_1 \tau_{0,0} \tau_{1,1} \bar{\tau}_{0,0} \pa_x \bar{\tau}_{3,1}
			+ \gamma_2 \tau_{1,1} \tau_{3,1} \bar{\tau}_{0,0}^2\right)
			\nonumber
			\\[0.01cm]
			+  \; \Gamma_2 &
			\left( \alpha_2 b \bar{r}_0 \bar{s}_0 \tau_{0,0} \tau_{3,1} \bar{\tau}_{0,0} \bar{\tau}_{1,1}
			+ \alpha_2 b s_0 \tau_{1,1} \pa_x \tau_{0,0} \bar{\tau}_{0,0} \bar{\tau}_{1,1} 	- \alpha_2 b s_0 \bar{r}_0 \bar{s}_0 \tau_{0,0} \tau_{1,1} \bar{\tau}_{0,0} \bar{\tau}_{1,1}
			\right. 
			\nonumber
			\\
			&\left.
			-  \; \alpha_2 b s_0 \tau_{0,0} \pa_x \tau_{1,1} \bar{\tau}_{0,0} \bar{\tau}_{1,1}
			+ \alpha_2 b s_0 \tau_{0,0} \tau_{1,1} \bar{\tau}_{1,1} \pa_x \bar{\tau}_{0,0}
			- \alpha_2 b s_0 \tau_{0,0} \tau_{1,1} \bar{\tau}_{0,0} \pa_x \bar{\tau}_{1,1}
			\right. 
			\nonumber
			\\
			&\left.
			+  \; b \gamma_1 \bar{s}_0 \tau_{0,0}^2 \bar{\tau}_{1,1}^2
			- \alpha_2 \tau_{3,1} \pa_x \tau_{0,0} \bar{\tau}_{0,0} \bar{\tau}_{1,1}
			+ \alpha_2 \tau_{0,0} \pa_x \tau_{3,1} \bar{\tau}_{0,0} \bar{\tau}_{1,1}
			\right. 
			\nonumber
			\\
			&\left.
			-  \; \alpha_2 \tau_{0,0} \tau_{3,1} \bar{\tau}_{1,1} \pa_x \bar{\tau}_{0,0}
			+ \alpha_2 \tau_{0,0} \tau_{3,1} \bar{\tau}_{0,0} \pa_x \bar{\tau}_{1,1}
			- \gamma_1 \tau_{0,0}^2 \bar{\tau}_{1,1} \bar{\tau}_{3,1} \right)=0,
		\end{align}
	\end{subequations}	
	
	where we defined the auxiliary fields $\Gamma_i$ that encodes vacuum information
	\begin{empheq}{align}
		\Gamma_1 = \exp(br_0 s_0 x)
		\qquad
		\text{and}
		\qquad
		\Gamma_2 = \exp(b\bar{r}_0 \bar{s}_0 x).
	\end{empheq}
	Finally, we recall that we must always satisfy  $\pa_x J_i - r_i s_i =0$ for $\text{i}=1,2$, which is equivalent to impose
	\begin{subequations}\label{J1.J2.tau}
		\begin{align}
			\mathcal{J}_1 &: \; \left(1-b\right) br_0s_0  \tau _{0,0} \tau _{1,1}+br_0 \tau _{0,0} \tau _{3,1}-bs_0 \tau _{1,1} \tau _{2,0}+\tau _{3,1} \tau _{2,0}-\tau _{1,1} \pa_x \tau _{0,0}+\tau _{0,0} \pa_x\tau_{1,1} = 0,
			\\[3mm]
			\mathcal{J}_2 &: \;  \left(1-b\right) b \bar{r}_0 \bar{s}_0 \bar{\tau }_{0,0} \bar{\tau }_{1,1} +b\bar{r}_0 \bar{\tau }_{0,0} \bar{\tau }_{3,1}-b \bar{s}_0 \bar{\tau }_{1,1} \bar{\tau }_{2,0}+\bar{\tau }_{3,1} \bar{\tau }_{2,0}-\bar{\tau }_{1,1}
			\pa_x\bar{\tau }_{0,0} + \bar{\tau }_{0,0} \pa_x \bar{\tau }_{1,1} = 0.
		\end{align}
	\end{subequations}
	
	In the next section, we will combine this set of equations with the previous solutions obtained in section \ref{sec.dressing}. As the tau are written as linear combinations of $\rho_i$, $\mathcal{B}_{\text{II}}$ this will lead to a set of polynomial equations for $\rho_i$ that are easier to solve than the original set of differential equations, allowing us to implement an efficient routine to study each case. Another advantage of such approach is that as we use the spacial part of the Lax pair to determine the Bäcklund transformation, the results here are valid for the \emph{entire} hierarchy.

	\subsection{Solitons and integrable defects}
	
	We have established the necessary framework to study different soliton solutions interacting with various integrable defects. We now provide different soliton solutions before and after the defect and require the Bäcklund transformations $\mathcal{B}_{\text{II}}$ to be satisfied. This will allow us to determine the final configuration of the  soliton  after interacting with the defect.
	
	The results are presented in the following order. First, we will consider the soliton solutions obtained via vertex operators, referred to as class A. We will perform this analysis for the following cases: one-soliton to one-soliton, one-soliton to two-soliton and two-soliton to two-soliton. Subsequently, we will apply the same procedure to the solitons obtained via vertex operators of the referred to as class B. Due to its complex structure, we only present the case of two-soliton to two-soliton transformation. For simplicity, in all cases we assume the same vacuum structure, i.e, $\bar{r}_0 = r_0$ and $\bar{s}_0 = s_0$. For all the cases, the solutions are written in terms of $	\rho_i$, such  for each integer flow N, we have:
	\begin{equation*}
		\rho_i 
		= e^{\left(k_i+br_0s_0\right) x +\omega_{i}^{\pm N} t} 
	\end{equation*}
	with
	\begin{align*}
		\omega_{i}^{N}=k_i^N- b \left(- r_{0}s_{0}\right)^N 
		\qquad 
		\text{or}  
		\qquad
		\omega_{i}^{-N}=\left( 1-2b\right)k_i^{-N}+ b\left(- r_{0}s_{0}\right)^{-N}.
	\end{align*}
	
	\subsubsection{One-soliton \texorpdfstring{$\rightarrow$}{->} one-soliton}
	
	Consider the one-soliton solution from class A \eqref{solA}, passing through the defect and emerging as another one-soliton class A shifted by a delay factor $R$:
	\begin{align}
		\begin{split}
			\tau_{0,0} &= 1 -b r_0 \rho_1,\\
			\tau_{1,1} &= 1,\\
			\tau_{2,0} &= br_0^2 \rho_1,\\
			\tau_{3,1} &= -k_1 \rho_1,
		\end{split}
		\begin{split}
			\bar{\tau}_{0,0} &= 1 - br_0 R \rho_1,\\
			\bar{\tau}_{1,1} &= 1,\\
			\bar{\tau}_{2,0} &= br_0^2 R \rho_1,\\
			\bar{\tau}_{3,1} &= -k_1 R \rho_1.
		\end{split}
	\end{align}
	We recall that class A solitons represent a natural reduction to the Burgers' equation \eqref{wpN0} as discussed before. The type II B\"acklund transformations are satisfied with the following conditions imposed on the delay factor ($R$) and Bäcklund parameters ($\alpha_i$, $\gamma_i$): 
	\begin{itemize}
		\item Scattering Conditions:
		\begin{itemize}
			\item $b=1$
			\begin{equation}
				R= \frac{k_1 \alpha_2+\g_1}{k_1 \alpha_1+\g_2} \quad \text{such} \quad \gamma _1 = \gamma _2+\left(\alpha _2-\alpha _1\right) r_0 s_0.
			\end{equation}
			\item $b=0$
			\begin{equation}
				R= \frac{k_1 \alpha_2+\g_1}{k_1 \alpha_2+\g_2}.
			\end{equation}
		\end{itemize}
		\item Solutions:
		\begin{equation}
			\left(r_1,s_1\right) = \left(br_0, \; \frac{bs_0+k_1 \rho_1}{1-br_0 \rho_1}\right)\quad \underset{\mathcal{B}^{\text{II}}}{\longrightarrow} \quad \left(r_2,s_2\right) =\left(br_0, \; \frac{bs_0+k_1 \; R \; \rho_1}{1-br_0 \; R \; \rho_1}\right).
		\end{equation}
	\end{itemize}
	Suitable limits as $\left(r_0,0\right)$ or $\left(0,s_0\right)$ can be considered for positive flows. After interacting with the defect, the one-soliton acquires a delay factor $R$. 
	
	\subsubsection{One-soliton \texorpdfstring{$\rightarrow$}{->} two-soliton}
	
	Now, we propose a one-soliton solution from the class A vertex operators, passing through the defect and emerging as another two-soliton class A. In terms of $\tau$-functions, we have
	\begin{align}
	\begin{split}
		\tau_{0,0} &= 1 -b r_0 \rho_1,\\
		\tau_{1,1} &= 1,\\
		\tau_{2,0} &= br_0^2 \rho_1 ,\\
		\tau_{3,1} &= -k_1 \rho_1,
	\end{split}
	\begin{split}
		\bar{\tau}_{0,0} &= 1 - br_0 (R \rho_1 +  \rho_2),\\
		\bar{\tau}_{1,1} &= 1,\\
		\bar{\tau}_{2,0} &=b r_0^2 (R \rho_1 +  \rho_2),\\
		\bar{\tau}_{3,1} &= -k_1 \; R \; \rho_1 - k_2 \; \rho_2,
	\end{split}
	\end{align}
	In this case, the defect converts a one-soliton configuration into a two-soliton configuration.
	We will have the following conditions upon the delay factor $R$ and Bäcklund parameters:
	\begin{itemize}
	\item Scattering Conditions:
	
	\begin{itemize}
		\item $b=1$
		\begin{equation}
			R=\frac{\gamma_1+ k_1 \al_2}{\gamma _2+k_1\alpha _1 } \quad \text{such} \quad \gamma _1 = \gamma _2-\left(\alpha _1-\alpha _2\right) r_0 s_0 \quad \text{and} \quad \g_2=-\al_1 k_2 .
		\end{equation}
		\item $b=0$
		\begin{equation}
			R= \frac{k_1 \alpha_2+\g_2}{k_1 \alpha_1+\g_1} \quad \text{and} \quad \g_1=-\al_1 k_2 .
		\end{equation}
	\end{itemize}
	\item Solutions:
	\begin{equation}
		\left(r_1,s_1\right) = \left(br_0, \; \frac{bs_0+k_1 \rho_1}{1-br_0 \rho_2}\right)\quad \underset{\mathcal{B}^{\text{II}}}{\longrightarrow} \quad \left(r_2,s_2\right) =\left(br_0, \; \frac{bs_0+k_1 R \rho_1+k_2\rho_2}{1-br_0 \;  \left(R\rho_1+\rho_2\right)}\right).
	\end{equation}
	\end{itemize}
	The resulting two-soliton emerges from the interactions of the one-soliton solution with the defect. It inherits the wave number $k_1$ and acquires a second wave number $k_2$.

	\subsubsection{Two-soliton \texorpdfstring{$\rightarrow$}{->} two-soliton}
	
	Now, we propose a two-soliton solution from the class A vertex operators, passing through the defect and emerging as another two-soliton class A, differing from the original by phases $R_1$ and $R_2$. In terms of $\tau$-functions, we have
	\begin{align}
	\begin{split}
		\tau_{0,0} &= 1 -b r_0 (\rho_1 + \rho_2),\\
		\tau_{1,1} &= 1,\\
		\tau_{2,0} &= b r_0^2 (\rho_1 + \rho_2),\\
		\tau_{3,1} &= -k_1 \rho_1-k_2 \rho_2,
	\end{split}
	\begin{split}
		\bar{\tau}_{0,0} &= 1 -b r_0 (R_1 \rho_1 + R_2 \rho_2),\\
		\bar{\tau}_{1,1} &= 1,\\
		\bar{\tau}_{2,0} &= br_0^2 (R_1 \rho_1 + R_2 \rho_2),\\
		\bar{\tau}_{3,1} &= -k_1 \; R_1 \; \rho_1 - k_2 \; R_2 \; \rho_2.
	\end{split}
	\end{align}
	
	The final result is given by:
	\begin{itemize}
	\item Scattering Conditions:
	\begin{itemize}
		\item $b=1$
		\begin{equation}
			R_i=\frac{\gamma_1+ k_i \al_2}{\gamma _2+k_i\alpha _1 } \quad \text{such} \quad \gamma _1 = \gamma _2-\left(\alpha _1-\alpha _2\right) r_0 s_0.
		\end{equation}
		\item $b=0$
		\begin{equation}
			R_i=\frac{\gamma_2+ k_i \al_2}{\gamma _1+k_i\alpha _1 }.
		\end{equation}
	\end{itemize}
	\item Solutions:
	\begin{equation}
		\left(r_1,s_1\right) = \left(br_0, \; \frac{bs_0+k_1 \rho_1+k_2 \rho_2}{1-br_0 \left(\rho_1+\rho_2\right)}\right)\quad \underset{\mathcal{B}^{\text{II}}}{\longrightarrow} \quad \left(r_2,s_2\right) =\left(br_0, \; \frac{bs_0+k_1 R_1 \rho_1+k_2 R_2 \rho_2}{1-br_0 \;  \left(R_1\rho_1+R_2\rho_2\right)}\right).
	\end{equation}
	\end{itemize}
	In this case two-soliton solution of class A interacting with defect, preserves the original waves number, $k_1$ and $k_2$ and acquires delay factors, $R_1$ and $R_2$.

	\subsubsection{Two-soliton \texorpdfstring{$\rightarrow$}{->} two-soliton}
	
	Now, we propose a two-soliton solution from the class B vertex operators, passing through the defect and emerging as another two-soliton class B, differing from the original by phase shifts $R_1$ and $R_2$:
	\begin{align}
	\begin{split}
		\tau_{0,0} &= 1-br_0 \rho_1 + \frac{k_2 \left(b r_0 s_0 + k_1 \right)^2}{\left(k_1-k_2\right)^2}\rho_1\rho _{2}^{-1},
		\\[2mm]
		\tau_{1,1} &= 1+b s_0 \rho _{2}^{-1}+ \frac{k_1 \left(br_0 s_0 + k_2\right)^2}{\left(k_1-k_2\right)^2}\rho_1\rho _{2}^{-1},
		\\[2mm]
		\tau_{2,0} &=b r_0^2 \rho_1 - k_2 \rho _{2}^{-1}- \frac{br_0 k_2 \left(k_1 + k_2 + 2 br_0 s_0 \right)}{(k_1 - k_2)} \rho_1 \rho _{2}^{-1},
		\\[2mm]
		\tau_{3,1} &=b s_0^2 \rho _{2}^{-1}- k_1 \rho_{1} - \frac{bs_0 k_1 \left(k_1 + k_2 + 2 br_0 s_0 \right)}{(k_1 - k_2)} \rho_1 \rho _{2}^{-1},
	\end{split}
	\end{align}
	and
	\begin{align}
	\begin{split}
		\bar{\tau}_{0,0} &= 1- b r_0 \; R_1 \; \rho_1 + \frac{k_2 \left(b r_0 s_0 + k_1 \right)^2}{\left(k_1-k_2\right)^2}  \; R_1 \; R_2 \; \rho_1\rho _{2}^{-1},
		\\[2mm]
		\bar{\tau}_{1,1} &= 1 + b s_0 \; R_2 \;\rho _{2}^{-1}+ \frac{k_1 \left(b r_0 s_0  + k_2\right)^2}{\left(k_1-k_2\right)^2} \; R_1 \; R_2 \; \rho_1\rho _{2}^{-1},
		\\[2mm]
		\bar{\tau}_{2,0} &= br_0^2 \; R_1 \; \rho_1 - k_2 \; R_2 \;\rho _{2}^{-1}- \frac{br_0  k_2 \left(k_1 + k_2 + 2 br_0 s_0 \right)}{(k_1 - k_2)} \; R_1 \; R_2 \; \rho_1 \rho _{2}^{-1},
		\\[2mm]
		\bar{\tau}_{3,1} &= bs_0^2 \; R_2 \; \rho _{2}^{-1}- k_1 \; R_1 \; \rho_{1} - \frac{bs_0 k_1 \left(k_1 + k_2 + 2b r_0 s_0  \right)}{(k_1 - k_2)} \; R_1 \; R_2 \; \rho_1 \rho _{2}^{-1}.
	\end{split}
	\end{align}
	This leads to the following results
	
	\begin{itemize}
	\item Scattering Conditions:
	\begin{itemize}
		\item $b=1$
		\begin{equation}
			\gamma _1 = \gamma _2-\left(\alpha _1-\alpha _2\right) r_0 s_0	, \quad R_1= \frac{k_1 \al_2+\g_1}{k_1 \al_1+\g_2} \quad \text{and} \quad R_2=\frac{k_2 \al_1+\g_2}{k_2 \al_2+\g_1}.
		\end{equation}
		\item $b=0$
		\begin{equation}
			R_1= \frac{k_1 \al_2+\g_2}{k_1 \al_1+\g_1} \quad \text{and} \quad R_2=\frac{k_2 \al_1+\g_1}{k_2 \al_2+\g_2}.
		\end{equation}
	\end{itemize}
	\item Solutions:
	\begin{subequations}
		\begin{align}
			r_1 &= \frac{\left(k_1-k_2\right){}^2 \left(br_0-k_2 \rho _{2}^{-1}\right)+k_2 \rho_{2}^{-1} \rho _1 r_0 \left(k_2+br_0 s_0\right){}^2}{k_1 \rho _{2}^{-1} \rho _1 \left(k_2+br_0 s_0\right){}^2+\left(k_1-k_2\right){}^2 \left(\rho _{2}^{-1} bs_0+1\right)},
			\\[3mm]
			s_1 &= \frac{k_1 \rho _1 \left(\rho _{2}^{-1} bs_0 \left(k_1+br_0s_0\right){}^2+\left(k_1-k_2\right){}^2\right)+\left(k_1-k_2\right){}^2b s_0}{\rho _1 \left(k_2 \rho _{2}^{-1} \left(k_1+br_0 s_0\right){}^2-\left(k_1-k_2\right){}^2b r_0\right)+\left(k_1-k_2\right){}^2},
		\end{align}
	\end{subequations}
	and
	\begin{subequations}
		\begin{align}
			r_2 &= \frac{\left(k_1-k_2\right){}^2 \left(br_0-k_2 R_2\rho _{2}^{-1}\right)+k_2 R_2R_1\rho _{2}^{-1} \rho _1 br_0 \left(k_2+br_0 s_0\right){}^2}{k_1 R_2R_1\rho _{2}^{-1} \rho _1 \left(k_2+br_0 s_0\right){}^2+\left(k_1-k_2\right){}^2 \left(R_2\rho _{2}^{-1}bs_0+1\right)},
			\\[3mm]
			s_2 &= \frac{k_1 R_1\rho _1 \left(R_2\rho _{2}^{-1} bs_0 \left(k_1+br_0 s_0\right){}^2+\left(k_1-k_2\right){}^2\right)+\left(k_1-k_2\right){}^2 bs_0}{R_1\rho _1 \left(k_2R_2 \rho _{2}^{-1} \left(k_1+br_0 s_0\right){}^2-\left(k_1-k_2\right){}^2b r_0\right)+\left(k_1-k_2\right){}^2}.
		\end{align}
	\end{subequations}
	In this case, the two-soliton interacting with the defect acquires delay factors $R_1$ and $R_2$.
	\end{itemize}
	
	\section{Discussion and further developments}
	\label{discussion}
	
	In this paper we propose a universal framework to deal with generalized integrable hierarchies in the sense that higher grading semi-simple elements (of grade $a\geq 1$) can be incorporated within the Riemann-Hilbert-Birkhoff decomposition.
	
	The framework extends the usual class of soliton solutions associated to zero vacuum solutions.  These, in turn define a centerless Heisenberg sub-algebra that include different types of boundary conditions. In fact, the construction of different Heisenberg sub-algebras classify the possible vacua of the soliton solutions.
	
	In particular, we have shown the existence of a novel class of constant non-zero vacuum solutions which are constructed from a one parameter dependent (deformed) Heisenberg sub-algebra.
	
	Explicit examples were found and discussed for the mKdV ($a=1$) \cite{gomes_nonvanishing_2012} and CLL ($a=2$) hierarchies, \cite{aratyn_generalized_2025}. Other interesting new examples, for $a>2$, follow the general pattern and are under investigation.
	
	%
	
	In particular for the CLL hierarchy, by a judicious choice of vertex operators, we have constructed a class of solutions in which one of the field remains constant. This provides, in a closed form, a systematic construction of soliton solutions for the entire underlying Burgers hierarchy. 
	
	Moreover the grading structure of the affine algebra provide a systematic construction of Bäcklund transformation in terms of graded group elements. The main idea is to consider graded gauge transformations to map different solutions of the same flow equation.
	
	Examples of such construction were successfully employed for the generalized $A_r$-mKdV hierarchies \cite{de_carvalho_ferreira_gauge_2021, de_carvalho_ferreira_generalized_2021}. Here we have determined the Bäcklund transformation for the entire CLL hierarchy. The reduction procedure yields the Bäcklund transformation for the Burgers hierarchy.
	
	Several Bäcklund solutions were explicitly worked out for the CLL and its underlined Burgers hierarchies.
	
	So far we have employed the principal gradation in our examples. An interesting construction will be to consider higher grading with mixed gradations, e.g., higher grading Yajima-Oikawa hierarchy (derivative Yajima-Oikawa).  
	
	An interesting pattern emerging naturally from our construction is the parameter dependence in the Lax operators in vacuum, (\ref{vac-pos}) and (\ref{vac-neg}). Notice that the grading added to the power of $r_0$ or $ s_0$ in each term in (\ref{vac-pos}) and (\ref{vac-neg}) is a constant. 
	
	This suggests a second loop described by $\zeta$ is a dimension of either $r_0 $ or $s_0$ and an effective grading can be defined to be
	\begin{align*}
	\tilde Q = Q + \zeta \frac {d}{d\zeta}, 
	\end{align*}            
	The idea of affine Lie algebra with two loops was introduced in \cite{aratyn_kac-moody_1991}.

	\section*{Acknowledgments}
	
	We thank the referee for the useful comments. JFG thank CNPq and FAPESP for support. YFA thanks FAPESP for financial support under grant \#2022/13584-0.  GVL thanks FAPESP for financial support under grant \#2024/16787-4.  TCS thanks FAPESP for financial support under grant \#2026/02077-0.

	\appendix

	\section{The \texorpdfstring{$sl(2)$}{sl(2)} affine algebra}
	\label{app.algebra}

	
	The affine  Kac-Moody algebra $\hat{\mathcal{G}}$  is an infinite extension of 
	a Lie algebra $\mathcal{G}$,  
	\begin{equation*}
	\hat{\mathcal{G}}=L(\mathcal{G})\oplus\mathbb{C}\hat{c}\oplus\mathbb{C}\hat{d}, 
	\end{equation*}
	where
	\begin{equation*}
	L(\mathcal{G})=\mathcal{G}\otimes\mathbb{C}\left[ \zeta, \zeta^{-1}\right]=\left\{ X \otimes \zeta^{n} \, | \,X \in \mathcal{G}, \, n\in \mathbb{Z}\right\}
	\end{equation*}
	is the loop algebra of  $\mathcal{G}$, where $\zeta \in \mathbb{C}$ is its spectral parameter \cite{kac_infinite-dimensional_1990}.  The central term  $\hat{c}$ commutes with all other generators and the spectral derivative  $\hat{d}=\zeta \frac{d}{d\zeta}$ measures the power of the parameter $\zeta$.
	
	Considering  $\mathcal{G}=A_1 \sim sl(2)$, the generators $\{h, E_{\alpha}, E_{-\alpha}\}$  obey the following commutation relations
	$\comm{h}{E_{\pm \alpha}}=\pm 2E_{\pm \alpha}$ and $\comm{E_{\alpha}}{E_{-\alpha}}=h$. The  corresponding affine 
	algebra $\hat{\mathcal{G}}=\hat{A}_1$ is obtained considering $L(\mathcal{G})=\{h^{(n)}=\zeta^n h, E_{ \alpha}^{(n)}=\zeta^n E_{\alpha}, E_{-\alpha}^{(n)}=\zeta^n E_{-\alpha} \}$ with the normalization $\alpha^2=2$. Thus, the affine Kac-Moody $\hat{\mathcal{G}}=\hat{A}_1 $  generators read  $\{h^{(n)},  E_{\alpha}^{(m)}, E_{-\alpha}^{(m)}, \hat{c}, \hat{d}\}$, and the commutation relations in the Chevalley basis are given by 
	\begin{equation}\label{eqs1}
	\begin{aligned}
		\comm{h^{(n)}}{h^{(m)}}&=2n\delta_{n+m,0}\hat{c}, \quad & \comm{h^{(n)}}{E_{\pm \alpha}^{(m)}}&=\pm 2 E_{\pm \alpha}^{(n+m)}, \quad &\comm{E_{ \alpha}^{(n)}}{E_{-\alpha}^{(m)}}&= h^{(n+m)} +n\delta_{n+m,0}\hat{c}, \\
		\comm{E_{\pm \alpha}^{(n)}}{E_{\pm \alpha}^{(m)}}&=0, \quad &\comm{\hat{c}}{T^{(n)}}&=0, \quad &\comm{\hat{d}}{T^{(n)} }&=n T^{(n)},
	\end{aligned}
	\end{equation}
	with  $n,m \in \mathbb{Z}$, where $T^{(n)} \in \{h^{(n)}=\zeta^n h, E_{ \alpha}^{(n)}=\zeta^n E_{\alpha}, E_{-\alpha}^{(n)}=\zeta^n E_{-\alpha} \}$. The $sl(2)$ loop algebra is obtained considering $\hat{c}=0$  into the commutation relations above.

	An algebra $\hat{\mathcal{G}}$ can be decomposed into graded subspaces  as follows:
	\begin{equation*}
	\hat{\mathcal{G}}=\bigoplus_{n \in \mathbb{Z} }\hat{\mathcal{G}}_n, \quad \quad \comm{\hat{\mathcal{G}}_{n}}{\hat{\mathcal{G}}_m}\subset \hat{\mathcal{G}}_{n+m}, \quad \quad  n,m \in \mathbb{Z},
	\end{equation*}
	where $\hat{\mathcal{G}}_n$ is a subspace of degree $n$ according to a grading operator
	$Q$ such that 
	\begin{equation*}
	\comm{Q}{{\hat{\mathcal{G}}}_n} = n \,\hat{\mathcal{G}}_n.
	\end{equation*}
	
	For $\hat{\mathcal{G}}=\hat{A}_1$  it is possible to define  two gradations, the homogeneous and the principal, whose grading operators are
	\begin{align*}
	Q_{h}=\hat{d}, \qquad Q_{p}=\frac{1}{2}h^{(0)}+2\hat{d}.
	\end{align*}
	
	Observe that in  order to construct the integrable hierarchies by an algebraic method it is enough to consider only the loop algebra, but once we want to obtain the solutions of the equations we need  to take into account  the complete Kac-Moody algebra and its representation theory. 
	
	The highest weight states of  $\hat{A}_1$, namely  $\ket{\mu_0}$ and $\ket{\mu_1}$, 
	are annihilated by all generators $T^{(n)}$  with $n>0$ and satisfy also:
	\begin{align*}
	\begin{split}
		h^{(0)}\ket{\mu_0}&=0,
		\\[2mm]
		E^{(0)}_{\alpha}\ket{\mu_0}&=0,
		\\[2mm]
		\hat{c}\ket{\mu_0}&=\ket{\mu_0},
	\end{split}
	\begin{split}
		h^{(0)}\ket{\mu_1}&=\ket{\mu_1},
		\\[2mm]
		E^{(0)}_{\alpha}\ket{\mu_1}&=0,
		\\[2mm]
		\hat{c}\ket{\mu_1}&=\ket{\mu_1}.
	\end{split}
	\end{align*}
	The adjoint relations read
	\begin{equation}
	\left( h^{(n)} \right)^{\dagger}=h^{(-n)}, \quad \left( E^{(n)}_{\pm \alpha} \right)^{\dagger}=E^{(-n)}_{\mp \alpha}, \quad \left( \hat{c} \right)^{\dagger}=\hat{c},
	\end{equation}
	thus, $\bra{\mu_0}$ and $\bra{\mu_1}$  are annihilated by all generators $T^{(n)}$  ($n<0$). 
	
	A $2 \times 2$ matrix representation  of the $sl(2)$  loop algebra is given by:
	\begin{align*}
	h^{(n)}= \left ( 
	\begin{array}{cc}
		\zeta^{n} & 0 \\
		0 & -\zeta^{n}
	\end{array}
	\right ),
	\qquad
	E_{\alpha}^{(n)}= \left ( 
	\begin{array}{cc}
		0 & \zeta^{n} \\
		0 & 0
	\end{array}
	\right ),
	\qquad
	E_{-\alpha}^{(n)}= \left ( 
	\begin{array}{cc}
		0 & 0 \\
		\zeta^{n} & 0
	\end{array}
	\right ).
	\end{align*}
	
	\section{Matrix Elements} \label{ap.matrix.el}
	\label{B}
	
	In this section we present the matrix elements used in order to obtain the solutions of Section \ref{sec.dressing}. In all cases we consider the vertices  (\ref{V+}) and (\ref{V-}) with their respective parameters $k_i \neq 0$. For those matrix elements which involve only one vertex
	($V_i^{+}$ or $V_i^{-}$) with its respective parameter $k_i$, we obtain:
	\begin{align*}
	\begin{split}
		\bra{\lambda_0}V_i^+\ket{\lambda_0}&=-br_{0},
		\\[2mm]
		\bra{\lambda_1}V_i^+\ket{\lambda_1}&=0,
		\\[2mm]
		\bra{\lambda_2}V_i^+\ket{\lambda_0}&=b r_0^2,
		\\[2mm]
		\bra{\lambda_3}V_i^+\ket{\lambda_1}&=-k_i,
	\end{split}
	\begin{split}
		\bra{\lambda_0}V_i^-\ket{\lambda_0}&=0,
		\\[2mm]
		\bra{\lambda_1}V_i^-\ket{\lambda_1}&=bs_{0},
		\\[2mm]
		\bra{\lambda_2}V_i^-\ket{\lambda_0}&=-k_i,
		\\[2mm]
		\bra{\lambda_3}V_i^-\ket{\lambda_1}&=b s_0^2.
	\end{split}
	\end{align*}
	All  matrix elements  $\bra{\lambda_{k}}\left(V_i^{\pm}\right)^n\ket{\lambda_{l}}$ with $n\geqslant2$ are zero for vertices (\ref{V+}) and (\ref{V-}).  The same holds for matrix elements which involve the product of two or more vertices of the same kind ($V_i^{+}$ or $V_i^{-}$), even when related to distinct parameters $k_i$ as:
	\begin{align*}
	\begin{split}
		\bra{\lambda_{k}}V_i^{\pm}V_j^{\pm}\ket{\lambda_{l}}=0,  
	\end{split}
	\begin{split}
		\bra{\lambda_{k}}V_i^{\pm}V_j^{\pm}V_m^{\pm}\ket{\lambda_{l}}=0.
	\end{split}
	\end{align*}
	Finally, for the product  $V_i^+$ with $V_j^-$ (or $V_i^-$ with $V_j^+$), the matrix elements are:
	\begin{align*}
	\begin{split}
		\bra{\lambda_0}V_i^+V_j^-\ket{\lambda_0}&=\frac{k_j\left(k_i+b  r_{0}s_{0}\right)^2}{\left(k_i-k_j\right)^2},
		\\[2.5mm]
		\bra{\lambda_1}V_i^+V_j^-\ket{\lambda_1}&=\frac{k_i\left(k_j+b  r_{0}s_{0}\right)^2}{\left(k_i-k_j\right)^2},
		\\[2.5mm]
		\bra{\lambda_2}V_i^+V_j^-\ket{\lambda_0}&=-\frac{ b r_{0} k_j\left(k_i +k_j+2 b r_{0}s_{0}\right)}{k_i-k_j},
		\\[2.5mm]
		\bra{\lambda_3}V_i^+V_j^-\ket{\lambda_1}&=-\frac{ b s_{0} k_i \left(k_i+k_j+2 b r_{0}s_{0}\right)}{k_i-k_j},	
	\end{split}
	\begin{split}
		\bra{\lambda_0}V_i^-V_j^+\ket{\lambda_0}&=\frac{k_i\left(k_j+b  r_{0}s_{0}\right)^2}{\left(k_i-k_j\right)^2},
		\\[2.5mm]
		\bra{\lambda_1}V_i^-V_j^+\ket{\lambda_1}&=\frac{k_j\left(k_i+b  r_{0}s_{0}\right)^2}{\left(k_i-k_j\right)^2},
		\\[2.5mm]
		\bra{\lambda_2}V_i^-V_j^+\ket{\lambda_0}&=\frac{ b r_{0} k_i \left(k_i +k_j+2 b r_{0}s_{0}\right)}{k_i-k_j}, 
		\\[2.5mm]
		\bra{\lambda_3}V_i^-V_j^+\ket{\lambda_1}&=\frac{ b s_{0} k_j \left(k_i+k_j+2 b r_{0}s_{0}\right)}{k_i-k_j}. 
	\end{split}
	\end{align*}
	

	\bibliographystyle{elsarticle-num}
	
	\bibliography{example}
	
\end{document}